\documentclass[]{revtex4}
\usepackage{amsmath}
\begin{document}
\def\a{\alpha}
\def\b{\beta}
\def\e{\epsilon}
\def\p{\partial}
\def\m{\mu}
\def\n{\nu}
\def\t{\tau}
\def\s{\sigma}
\def\g{\gamma}
\def\r{\rho}
\def\half{\frac{1}{2}}
\def\hatt{{\hat t}}
\def\hatx{{\hat x}}
\def\hatp{{\hat p}}
\def\hatX{{\hat X}}
\def\hatY{{\hat Y}}
\def\hatP{{\hat P}}
\def\hatth{{\hat \theta}}
\def\hatta{{\hat \tau}}
\def\hatrh{{\hat \rho}}
\def\hatva{{\hat \varphi}}
\def\barx{{\bar x}}
\def\bary{{\bar y}}
\def\barz{{\bar z}}
\def\p{\partial}
\def\nn{\nonumber}
\def\cb{{\cal B}}
\def\2pap{2\pi\alpha^\prime}
\def\wideA{\widehat{A}}
\def\wideF{\widehat{F}}
\def\beq{\begin{eqnarray}}
 \def\eeq{\end{eqnarray}}
 \def\2pap{2\pi\a^\prime}
 \def\xp{x^\prime}
 \def\xpp{x^{\prime\prime}}
 \def\xppp{x^{\prime\prime\prime}}
 \def\barxp{{\bar x}^\prime}
 \def\barxpp{{\bar x}^{\prime\prime}}
 \def\barxppp{{\bar x}^{\prime\prime\prime}}
 \def\barchi{{\bar \chi}}
 \def\bpsi{{\bar \psi}}
 \def\barg{{\bar g}}
 \def\barz{{\bar z}}
 \def\bareta{{\bar \eta}}
 \def\ta{{\tilde a}}
 \def\tb{{\tilde b}}
 \def\tc{{\tilde c}}
 \def\tpsi{\tilde{\psi}}
 \def\tal{{\tilde \alpha}}
 \def\tbe{{\tilde \beta}}
 \def\barth{{\bar \theta}}
 \def\bareta{{\bar \eta}}
 \def\barom{{\bar \omega}}

 \title[Short Title]{Fermion Representation Of \\ The Rolling Tachyon
 Boundary Conformal Field Theory}

 \author{Taejin Lee}
 \email{taejin@kangwon.ac.kr}
 \author{Gordon W. Semenoff}
 \email{semenoff@nbi.dk}

 \affiliation{$^*$Department of Physics, Kangwon National University,
 Chuncheon 200-701, Korea}

 \affiliation{$^*$Asia Pacific Center for Theoretical Physics, Pohang 790-784,
 Korea}

 \affiliation{
 $^{*\dagger}$Department of Physics and Astronomy
 University of British Columbia \\6224, Agricultural Road
 Vancouver, B.C. V6T 1Z1
 Canada}

 \date{\today}

 \begin{abstract}
 A free fermion representation of the rolling tachyon boundary
 conformal field theory is constructed.  The representation is used to
 obtain an explicit, compact, exact expression for the boundary state.
 We use the boundary state to compute the disc and  cylinder amplitudes for
 the half-S-brane.

 \end{abstract}

 \maketitle

 \vskip .75cm \section{Introduction}

 One of the most important puzzles in string theory is the fate of
 unstable D-branes and how to describe their time evolution toward
 that fate. An intriguing conjecture for their time evolution is due
 to Ashoke Sen \cite{sen02r} and is called the rolling tachyon 
 \cite{senreview}. It asserts the existence of an exact time-dependent 
 classical solution of string theory describing the decay of a D-brane by the
 condensation of open string tachyons.

In the sigma model approach to the boson string, the rolling tachyon
is obtained by adding an exactly marginal boundary operator to the
world-sheet action, \beq \label{tachyonaction}S = -\frac{1}{4\pi}
\int_M d\tau d\sigma \p X^0\cdot \p X^0 + \oint_{\p M} d\s
\left(Ae^{X^0}+Be^{-X^0}\right) +\ldots\eeq We use units where
$\a^\prime =1$.  The second term is the boundary interaction
corresponding to the tachyon field, with $A$ and $B$ constants
determined by the initial conditions. The remaining terms denoted by
three dots are the action of a $c=25$ conformal field theory and
ghosts.

In the idealized situation, where the tachyon condensate is
space-independent, the other degrees of freedom decouple from the
time component $X^0$ and we can study it by itself.  We shall assume
that this is the case, and will not discuss the other coordinates or
ghosts further in this Paper.  Note the negative signature of the
kinetic term for $X^0$ arising from the fact that it is the time
coordinate.

 The simplest choice for the tachyon profile  in (\ref{tachyonaction})
is $A=\frac{g}{2}$ and $B=0$. With this profile the conformal
field theory (\ref{tachyonaction}) describes an unstable D-brane
in the past which then decays into a more stable configuration. In
this Paper, we will focus on this configuration, sometimes called
the half-S-brane.  The profile originally discussed by Sen
\cite{sen02r} has $A=B=g/2$ is termed the full-S-brane.
\cite{lambert}.

 To analyze this theory, we begin with a Wick rotation
 to Euclidean time, $X^0\to   iX^0$.
 After Wick rotation, the interaction in
 (\ref{tachyonaction}) is a periodic function of $X^0$.
 (From now on we will drop the
 superscript 0.) It describes a non-compact boson with a periodic boundary potential.

 The
 conformal field theory of a massless scalar field living on a strip
 with an exactly marginal periodic boundary interaction (with
 $A=B$) at one boundary and Dirichlet condition at the other boundary was discussed
 a decade ago in refs.\cite{call93,call94}.  In ref.\cite{call94} it
 was shown that the boundary state can be given as an exact sum of
 reparametrization invariant Ishibashi states \cite{ishi,cardy} of the
 $SU(2)$ current algebra.

Sen observed that
 this boundary state can be used to describe the rolling tachyon
 \cite{sen02r}.
 Since then, a large body of work has explored the implications of
 this exact, time dependent background
 \cite{sen02r,lambert,sen022,mukhopa,sen023,sen0305,sen031,sen0306,sen0308,larsen02,gibbons,sen02t,okuda,kim,hlee,sugino,
 ishida,buchel,chen,rey,rey2,taka,doug,arefeva,leblond,demasure,gutperle02,strominger02,gutperle03,larsen,constable,schomerus,nagami,foto}.
 It has greatly improved our understanding of the time dependent
 processes driven by the tachyon instability in string theory.
 However, our understanding of the final fate of the unstable D-brane
 and details of the dynamics of the rolling tachyon process is still far from
 complete. The exact time dependent description is only available for
 the first few lower levels and a complete description of the time
 evolution of the boundary state is still lacking.

 In this Paper we shall discuss a free fermion representation of the
boundary conformal field theory of the rolling tachyon.  Our
representation is a generalization of one which was proposed by
Polchinski and Thorlacius \cite{pol}. Like ref.\cite{call93,call94},
they studied a two-dimensional scalar conformal field theory on a
strip with a marginal periodic interaction at one boundary and
Dirichlet condition at the other boundary.  In order to find a free
fermion representation of the conformal field theory, they had to
introduce an extra boson.  The advantage of the resulting fermionic
representation is that the boundary interaction becomes a simple
fermion current operator which is bilinear in fermion fields. The
 result is a free field theory of fermions interacting with a
 boundary potential.  Ref.~\cite{pol} used it to compute the boundary
 S-matrix and the partition function which they found to agree with
 results of the current algebra technique \cite{call94}.

 Here we shall extend the fermion conformal field theory formulated in
 ref.\cite{pol} for the open string to the closed string sector.  We
will  discuss the construction of Neumann, Dirichlet and rolling
tachyon
 boundary states in the fermion language. We will find simple closed form
 expressions for them. In order to demonstrate the equivalence of the
 fermion and boson approaches, we use the fermion boundary state to
 compute the disc amplitude and find
 agreement with known results.  We also use it to compute the
 cylinder amplitude.

 One interesting feature of the closed string sector is
 that, to obtain all of the discrete momentum and wrapped
 sectors of the
 periodically identified boson, we must consider states from
 the Hilbert spaces of both Ramond
 and Neveu Schwarz fermions.

 \vskip .75cm \section{Preliminaries}

 Before we go on  to discuss the fermion representation of the
 rolling tachyon in later Sections, we pause to review some of the
 essentials of bosonization and fermionization in this Section.  We
 will discuss the representation of a closed string boson variable
 by fermions.  We will show how to find boundary
 states in terms of fermion variables and we will compute partition
 functions for closed and open strings.  This is a warmup exercise for
 the slightly  more complicated fermionization of the
 rolling tachyon system which will appear in later Sections.

 \subsection{Bosonization and Fermionization}

 In two dimensional conformal field theory, fermions and bosons are mapped to
 each other by
 \begin{subequations}
 \label{generallabel}
 \begin{eqnarray}
 \psi_L(z)~=~e^{-i\frac{\pi}{2}p_R}~:e^{-i\sqrt{2}\phi_L(z)}:
 ,~~~
 \psi_L^{\dagger}(z)~=~e^{i\frac{\pi}{2}p_R}~:e^{i\sqrt{2}\phi_L(z)}:
 \label{bosform:a}\\
 \psi_R(\bar z)~=~e^{-i\frac{\pi}{2}p_L}~:e^{i\sqrt{2}\phi_R(\bar
 z)}: ,~~~ \psi_R^{\dagger}(\bar
 z)~=~e^{i\frac{\pi}{2}p_L}~:e^{-i\sqrt{2}\phi_R(\bar z)}:
 \label{bosform:b}
 \end{eqnarray}
 \end{subequations}
 where  $\phi_R(\bar z)$ and $\phi_L(z)$ are the right- and
 left-moving boson fields, respectively.  The exponentials $e^{\pm
 i\frac{\pi}{2}p_{L,R}}$ are cocycles, necessary to make $\psi_R(\bar
 z)$ and $\psi_L(z)$ anti-commute. (See Appendix for more about
 cocycles.) Here $p_L$ and $p_R$ are moment and the normal ordering
 is defined below.

The complex number $z$ is the coordinate of the complex plane.  The
coordinates of the Euclidean cylinder, $(\tau,\sigma)$,
are related to it by the conformal map $z=e^{\tau+i\sigma}$,
$\bar z=e^{\tau-i\sigma}$.
The left- and right-moving boson
operators are defined by the mode expansions:
 \begin{subequations}
 \label{generallabel}
 \begin{eqnarray}
 \phi_L(\tau+i\sigma)~=~
 \frac{1}{\sqrt{2}} x_L-\frac{i}{\sqrt{2}} p_L(\tau+i\sigma)+\frac{i}{\sqrt{2}}\sum_{n\neq
 0}\frac{\beta_n}{n}e^{-n(\tau+i\sigma)}, \label{expan:a} \\
 \phi_R(\tau-i\sigma)~=~
 \frac{1}{\sqrt{2}} x_R- \frac{i}{\sqrt{2}} p_R(\tau-i\sigma)+\frac{i}{\sqrt{2}}\sum_{n\neq
 0}\frac{\tilde\beta_n}{n}e^{-n(\tau-i\sigma)}. \label{expan:b}
 \end{eqnarray}
 \end{subequations}
  The non-vanishing commutators are
 \begin{subequations}
 \label{generallabel}
 \begin{eqnarray}
 \left[ x_L,p_L\right]~&=&~i ~~~,~~~ \left[ x_R, p_R\right] ~=~ i
 \label{commu:a}\\
 \left[ \beta_m, \beta_n \right] ~&=&~ m\delta_{m+n} ~~~,~~~ \left[
 \tilde\beta_m, \tilde\beta_n \right] ~=~ m\delta_{m+n}.
 \label{commu:b}
 \end{eqnarray}
 \end{subequations}
 A representation of this algebra begins with a vacuum $|p_L,p_R\rangle$
 which is an eigenstate of momenta $(p_L,p_R)$ and is annihilated by
 positively moded oscillators
 \begin{equation}
 \beta_k |p_L,p_R\rangle= 0
  ~~,~~ \tilde \beta_k |p_L,p_R\rangle=0~,~~k>0
   ~.\end{equation}
 Excited states are created by negatively moded oscillators,
 $$
 (\beta_{-1})^{k_1}(\beta_{-2})^{k_2}...
 (\tilde\beta_{-1})^{\bar k_1}(\tilde\beta_{-2})^{\bar k_2}...|p_L,p_R\rangle~~~,~k_n,\bar k_n=0,1,2,...
 $$
 Normal ordering of any operator puts all negatively moded
 operators to the left of positively moded
 operators.

 The equal-time commutation relation in Euclidean space-time is
 \begin{equation}
 \left[ \phi(\tau,\sigma), \partial_\tau \phi(\tau, \sigma')\right]
 =2\pi\delta(\sigma-\sigma')
 \end{equation}
 where
 \begin{equation}
 \phi(\tau,\sigma)=\phi_L(\tau+i\sigma)+\phi_R(\tau-i\sigma)
 \label{lrdecomp}.
 \end{equation} 
 The commutation relations and equation of motion
 $$
 \left(\frac{\partial^2}{\partial\tau^2}+\frac{\partial^2}{\partial\sigma^2}\right)\phi=0
 $$
 which is solved by the decomposition
 (\ref{lrdecomp}) are obtained by canonical quantization of the field
 theory with action
 \begin{equation} S=\frac{1}{4\pi}\int d\tau d\sigma~
 \partial_a \phi~ \partial_a \phi. \end{equation}

 Fermions have the mode expansion \begin{subequations}
 \label{generallabel} \begin{eqnarray} \psi_L(\tau+i\sigma)= \sum_n
 \psi_n e^{-n(\tau+i\sigma)} ~~,~~
 \psi^{\dagger}_L(\tau+i\sigma)=\sum_n
 \psi_n^{\dagger}e^{-n(\tau+i\sigma)} \label{fermode:a}\\
 \psi_R(\tau-i\sigma)=\sum_n \tilde \psi_n e^{-n(\tau-i\sigma)} ~~,~~
 \psi^{\dagger}_R(\tau-i\sigma) = \sum_n \tilde \psi_n^{\dagger}
 e^{-n(\tau-i\sigma)}\label{fermode:b} \end{eqnarray}
 \end{subequations} with non-vanishing anticommutation relations
 \begin{equation} \label{anticomns} \left\{\psi_m, \psi_n^{\dagger}
 \right\} = \delta_{m+n} ~~~,~~~ \left\{ \tilde \psi_m , \tilde
 \psi_n^{\dagger}\right\}=\delta_{m+n}. \end{equation} We must
 distinguish between the cases when the fermions have periodic or
 anti-periodic wave-functions on the cylinder.  When they are
 anti-periodic their mode expansion has oscillators $\psi_n$ labeled
 by $n$ which are half-odd-integers, $(n\in {\bf Z} +1/2)$ so that
 $(\psi_{L},\psi_R)\to (-\psi_{L},-\psi_R)$ when
 $\sigma\to\sigma+2\pi$.  The oscillators can be uniquely separated
 into those with positive and negative modes.  A representation of
 (\ref{anticomns}) is found by beginning with the vacuum state which
 is annihilated by all positively moded oscillators \begin{equation}
 \left.\begin{matrix} \psi_n \left| 0\right\rangle_{NS}=0,\,\, & \tilde
 \psi_n\left|0\right\rangle_{NS}=0 \cr \psi^{\dagger}_n \left|
 0\right\rangle_{NS}=0,\,\, & \tilde \psi^{\dagger}_n\left|0\right\rangle_{NS}=0
 \cr \end{matrix} \right\}~~n>0 \end{equation} and creating excited
 states by operating on the vacuum with negatively moded oscillators
  $$ ( \psi_{ -\frac{1}{2} } )^{ k_{1/2} }
 (\psi_{-\frac{3}{2}})^{k_{3/2}}...
 ( \tilde\psi_{ -\frac{1}{2} } )^{ k_{1/2}' }
 (\tilde\psi_{-\frac{3}{2}})^{k_{3/2}'}...
 ( \psi^{\dagger}_{ -\frac{1}{2} } )^{ \bar k_{1/2} }
 ( \psi^{\dagger}_{ -\frac{3}{2} } )^{\bar k_{3/2} }...
 ( \tilde\psi^{\dagger}_{ -\frac{1}{2} } )^{ \bar k_{1/2}' }
 ( \tilde\psi^{\dagger}_{ -\frac{3}{2} } )^{\bar k_{3/2}' }...
 |0\rangle $$ $${\rm with}~~~~~k_n,\bar k_n,k_n',\bar k_n'=0,1.
 $$
 The fermion theory with anti-periodic boundary conditions and
 half-integrally moded oscillators is called the Neveu-Schwarz (NS)
 sector.  When both left- and right-moving fermions are antiperiodic it
 is called the NS-NS sector.

 When the boundary condition for the fermions is periodic, the
 oscillators are labeled by integers $(n\in{\bf Z})$. Then, the
 oscillator with $n=0$ is a fermion zero mode.  Because of the
 presence of this zero mode, the states are degenerate. We begin with
 the state $|--\rangle$ defined by
 \begin{equation}
 \left.\begin{matrix} \psi_n \left|--\right\rangle=0~, & \tilde \psi_n\left|--\right\rangle=0
 & ~~n\geq0
 \cr  \psi^{\dagger}_n \left| --\right\rangle=0~, & \tilde
 \psi^{\dagger}_n\left|--\right\rangle=0 & ~~n > 0 \cr
 \end{matrix} \right\} .
 \end{equation}
 Then, the  degenerate vacuum states are $\left|++\right\rangle=
 \psi_0^{\dagger}\tilde \psi_0^{\dagger}\left|--\right\rangle$,
 $\left|+-\right\rangle=\psi_0^{\dagger}\left|--\right\rangle$,
 $\left|-+\right\rangle=\tilde \psi_0^{\dagger}\left|--\right\rangle$,
 $\left|--\right\rangle$. Excited states are created from the four vacuum
 states by operating with negatively moded oscillators.

 A fermion with periodic boundary conditions is called a Ramond
(R) fermion and the theory which we are discussing where both left-
and right-moving fermions are periodic is called the R-R sector.

 The non-vanishing equal-time anti-commutation relations are
  \begin{eqnarray}
  \left\{ \psi_L(\tau,\sigma),
  \psi^{\dagger}_L(\tau,\sigma')\right\}
  =2\pi\delta(\sigma-\sigma')
  ~~,~~
  \left\{ \psi_R(\tau,\sigma),
  \psi^{\dagger}_R(\tau,\sigma')\right\}
  =2\pi\delta(\sigma-\sigma').
 \end{eqnarray}
 The fermions with this anti-commutation relation and the equations of
 motion
 $$
 \left(\partial_\tau+i\partial_\sigma\right)\psi_L=0
 ~~,~~\left(\partial_\tau-i\partial_\sigma\right)\psi_R=0
 $$
 result from quantization of the field theory with Euclidean action
 \begin{equation}
 S=\frac{1}{2\pi}\int d\tau d\sigma ~\left[ \psi_L^{\dagger}\left(
 \partial_\tau+ i\partial_\sigma\right)\psi_L +
 \psi_R^{\dagger}\left(\partial_\tau-
 i\partial_\sigma\right)\psi_R\right]. \label{fermstringaction}
 \end{equation}
We shall see in the following that we shall need to use
 both R-R and NS-NS fermions.

The fermion theory produces states where the momenta $p_L$ and $p_R$
have certain discrete values. To see this, we recall that  momenta
are related to fermion numbers. The number operators are obtained by
integrating
 \begin{equation}
 :\psi^{\dagger}_L(z)\psi_L(z):~=~i\sqrt{2}\partial_\tau \phi_L(z)
 ~~,~~ :\psi^{\dagger}_R(\bar z)\psi_R(\bar
 z):~=~-i\sqrt{2}\partial_\tau \phi_R(\bar z) \label{bosform:c}
 \end{equation}
 which can be gotten from Eqs.~(\ref{bosform:a}), (\ref{bosform:b})
 and the operator product expansion.  The momenta are then
 \begin{equation}\label{fermnum}
 p_L~=~\int_0^{2\pi}\frac{d\sigma}{2\pi}:\psi_L^{\dagger}(z)\psi_L(z):
 ~~,~~ p_R~=~-\int_0^{2\pi}\frac{d\sigma}{2\pi}:\psi_R^{\dagger}(\bar
 z)\psi_R(\bar z):.
 \end{equation}
 In our quantization of the fermions,
 the fermion number operators have integer spectra in the NS sector
 and, because of the fermion zero modes, half-odd-integer spectra in
 the R sector. In the NS-NS-sector

 \begin{equation}\label{fermnumns}
 p_L~=~\sum_{n=\half}^\infty\left(\psi^{\dagger}_{-n}\psi_{n}-\psi_{-n}\psi^{\dagger}_n\right)
 ~~,~~
 p_R~=~-\sum_{n=\half}^\infty\left(  {\tilde\psi}^{\dagger}_{-n}{\tilde\psi}_{n}-
 {\tilde\psi}_{-n}{\tilde\psi}^{\dagger}_n  \right)
 \end{equation}
and in the R-R sector
 \begin{equation}\label{fermnumr}
 p_L=\sum_{n=1}^\infty
\left( \psi^{\dagger}_{-n}\psi_{n}-\psi_{-n}\psi^{\dagger}_n \right)
 +\psi_0^{\dagger}\psi_0-\frac{1}{2}
 ~~,~~
 p_R=-\sum_{n=1}^\infty
\left( {\tilde\psi}^{\dagger}_{-n}{\tilde\psi}_{n}-
 {\tilde\psi}_{-n} {\tilde\psi}^{\dagger}_n \right)
 -{\tilde\psi}^{\dagger}_0{\tilde\psi}_0+\frac{1}{2}.
 \end{equation}
Consistent with this, in the formula
(\ref{bosform:a}), for example, the momentum and coordinate appear as
 \begin{equation}\label{rotate}\psi_L=e^{-i\frac{\pi}{2}p_L}
 e^{-ix_L-p_L\tau-ip_L\sigma+\ldots}. 
 \end{equation}
 Then, unraveling operator ordering gives  
 $$\psi_L(\tau,\sigma+2\pi)=-e^{-2\pi i
 p_L}\psi_L(\tau,\sigma).$$ 
 From this, we see that, to get anti-periodic NS fermions, the
 momentum $p_L$ should be quantized as integers, whereas to get
 periodic R fermions, it should be quantized as half-odd-integers.

The fact that the fermion theory produces only states where the
momenta are quantized in either integer or half-odd-integer units
means that they can only correspond to some of the states of the
boson theory where, for a non-compact boson, the momentum is a
continuously varying quantum number.  On the other hand, this
changes when the boson is periodically identified.  Then, when the
spatial coordinate of the boson is periodic, the momentum is
discrete, and we might hope to match the discrete momenta of
compact boson with fermion numbers in the fermion theory. Indeed,
the periodicity of the boson should be compatible with the
bosonization formulae, (\ref{bosform:a}) and (\ref{bosform:b}).

We can see this more clearly by comparing partition functions of the
two theories.
 The Hamiltonian in the boson and fermion representations are
 \begin{subequations}
 \label{generallabel}
 \begin{eqnarray}
 \label{ham1} L_0^B &=& \frac{1}{2}p_L^2+\sum_{n=1}^\infty
 \beta_{-n}\beta_n- \frac{1}{24} ~~,~~ \tilde
 L_0^B=\frac{1}{2}p_R^2+\sum_{n=1}^\infty\tilde\beta_{-n}\tilde\beta_n-
 \frac{1}{24},\\ \label{ham2} 
 L^{NS}_0 &=& \sum_{n\in{\bf Z}+1/2}n:
 \psi_{-n}^\dagger \psi_n: -\frac{1}{24} ~~,~~ \tilde
 L^{NS}_0 = \sum_{n\in{\bf Z}+1/2} n:\tilde \psi_{-n}^\dagger \tilde
 \psi_n:-\frac{1}{24},\\ \label{ham3} 
 L^R_0 &=& \sum_{n\in{\bf Z}} n:
 \psi_{-n}^\dagger \psi_n: +\frac{1}{12} ~~,~~ 
 \tilde L^R_0 = \sum_{n\in{\bf Z}} n:\tilde \psi_{-n}^\dagger \tilde
 \psi_n:+\frac{1}{12} 
 \end{eqnarray}
 \end{subequations} where, in (\ref{ham1}),
 $\beta_0=p_L$ and $\tilde\beta_0=p_R$.  
 The partition functions for left-movers are
 \begin{subequations}
 \label{generallabel}
 \begin{eqnarray} 
 {\rm Tr}\left[q^{L_{0}^{B}}\right]~&=&~q^{-\frac{1}{24}}\sum_{p_L}
 q^{\frac{1}{2}p_L^2}~\prod_{n=1}^\infty \frac{1}{1-q^n} 
 \label{part:1} \\ 
 {\rm Tr}\left[q^{L_{0}^{NS}}\right] &=& q^{-\frac{1}{24}}~\prod_{n=0}^\infty
 (1+q^{n+\frac{1}{2}})^2 \label{part:2}\\ 
 {\rm Tr}\left[q^{L_{0}^{R}}\right]
 ~&=&~q^{\frac{1}{12}}~2\prod_{n=1}^\infty (1+q^n)^2. \label{part:3} 
 \end{eqnarray}
 \end{subequations}
 Using the Jacobi triple product identity 
 \beq \label{jacobi}
 \sum_{n\in Z} z^n q^{n^2} = \prod_{n=0}^\infty (1-q^{2n+2})(1+zq^{2n+1})
 (1+z^{-1}q^{2n+1}),
  \eeq 
 we can derive the identities
 \begin{subequations}
 \label{generallabel}
 \begin{eqnarray} q^{-\frac{1}{24}}~\sum_{n\in{\cal Z}}
 q^{\frac{1}{2}n^2}~\prod_{n=1}^\infty \frac{1}{1-q^n} ~&=&~
 q^{-\frac{1}{24}}\prod_{n=0}^\infty (1+q^{n+ \frac{1}{2} } )^2
 \label{id1} \\ q^{-\frac{1}{24}}~\sum_{n\in{\cal Z}}
 q^{\frac{1}{2}(n+\frac{1}{2})^2}~\prod_{n=1}^\infty
 \frac{1}{1-q^n}~&=&~q^{\frac{1}{12}}~2\prod_{n=1}^\infty(1+q^n)^2
\label{id2} \\
 q^{-\frac{1}{24}}~\sum_{n\in{\cal Z}}  (-1)^n
 q^{\frac{1}{2}n^2}~\prod_{n=1}^\infty \frac{1}{1-q^n} ~&=&~
 q^{-\frac{1}{24}}\prod_{n=0}^\infty (1-q^{n+ \frac{1}{2} } )^2
 .\label{id3}
\end{eqnarray}
\end{subequations}

 Using these identities, we see that boson partition function with
 $p_L$ restricted in various ways is equivalent to various fermion
 partition functions:
 \begin{subequations}
 \label{generallabel}
 \begin{eqnarray}
 {\rm Tr}\left[q^{L_{0}^{B}}\right]={\rm Tr}\left[q^{L_{0}^{NS}}\right]~~~&&{\rm
 when}~p_L=\frac{1}{2}\cdot\left({\rm ~even~integer}~\right)
 \\
 {\rm Tr}\left[q^{L_{0}^{B}}\right]= {\rm Tr}\left[q^{L_{0}^{R}}\right] ~~~&&{\rm when
 }~p_L=\frac{1}{2}\cdot({\rm ~odd~ integer})  \\  {\rm
 Tr}\left[q^{L_{0}^{B}}\right]={\rm Tr}\left[q^{L_{0}^{NS}}\right]~
 +~{\rm Tr}\left[q^{L_{0}^{R}}\right]
 ~~~&&{\rm when }~p_L=\frac{1}{2}\cdot({\rm  ~ integer}~).
 \end{eqnarray}
 \end{subequations}
 A similar consideration holds for the right-moving sector.

 It is interesting to ask how we could reproduce the partition
 function of a compact boson field. Consider
the periodic identification
 \begin{equation}\label{periodic}
 \phi ~\sim ~\phi~+~2\pi R
 \end{equation}
 Because the coordinate is identified periodically, the
 total momentum is quantized.  In addition, the string world-sheet can wrap the
 target space cycle, so there is a quantized wrapping number.
 The quantized total momentum is the sum of left- and
 right-moving momenta. The wrapping number is proportional to
 the difference of left- and
 right-moving momenta, which is therefore also quantized.
 The result is
 \begin{equation}
 p_L+p_R= \frac{\sqrt{2}}{R}\cdot{\rm ~integer} ~~,~~ p_L-p_R =
 \sqrt{2}R\cdot{\rm ~integer}.
 \label{compact1}\end{equation}
 This correlates the momenta in the right and left-handed sectors.
 The appropriate identifications are
 \begin{equation}\label{compact2}
 R=\frac{1}{\sqrt{2}}
 \end{equation}
 or its T-dual $R=\sqrt{2}$.  Note that both of these would be consistent with the
 bosonization formulae (\ref{bosform:a}) and (\ref{bosform:b}).  (When $R=1/\sqrt{2}$ the
 right and left-handed fermions can be double-valued.)  We will consider $R=1/\sqrt{2}$.
 Then
 \begin{equation}
 p_L+p_R=2\cdot{~\rm integer} ~~,~~ p_L-p_R=~{\rm integer}
 \label{momentumintegers}\end{equation}
This determines the spectra of the momentum operators $p_L$ and
$p_R$. Clearly, these equations are solved when both $p_L$ and $p_R$
are integers, or when both are half-odd-integers, with the
additional constraint that their sum is an even integer.

  To fermionize this system, we
 must combine different sectors of fermions in such a way as to
 reproduce this spectrum of momenta.  It is straightforward to
 show that
 \begin{eqnarray}\label{gso}
 {\rm Tr}\left[q^{L_{0}^{B}} {\bar q}^{\tilde L_{0}^{B}}\right] =
 {\rm Tr}\left[\frac{1}{2}
 \left(1+(-1)^{p_L+p_R}\right)q^{L_{0}^{NS}}~{\bar q}^{\tilde
 L_{0}^{NS}}\right]~+~{\rm Tr}\left[\frac{1}{2}
 \left(1+(-1)^{p_L+p_R}\right)q^{L_{0}^{R}}~{\bar q}^{\tilde
 L_{0}^{R}}\right].
 \end{eqnarray}
 In the fermion traces, $p_L$ and $-p_R$ are identified with
 the fermion number operators
 (\ref{fermnum}). Remember that $p_L$ and $p_R$ are integers in the NS-NS
 sector and are half-odd-integers in the R-R sector.

 In Eq.~(\ref{gso}), we have enforced the condition that $p_L+p_R$ must be an
 even integer by inserting the projection operator  $\half\left(1+(-1)^{p_L+p_R}\right)$.
 The trace in the NS-NS sector sums over those states where both $p_L$ and $p_R$ are integers
 and the sum $p_L+p_R$ is even, therefore the difference $p_L-p_R$ is  also
 even. The trace in the R-R sector sums
 over states where both $p_L$ and $p_R$ are half-odd-integers, then,
 when $p_L+p_R$ is even, $p_L-p_R$ is  odd.
 Including both sectors then sums over
 states where $p_L+p_R$ is even and $p_L-p_R$ are any integers, matching the spectrum
 of the compact boson with identification $X\sim X+\sqrt{2}\pi$ exactly.

The projection in Eq.~(\ref{gso}) is reminiscent of the GSO
projection of the fermionic  Neveu-Schwarz-Ramond string which
obtains the non-supersymmetric type 0 string theory.  In the
following, we shall see that different compactifications of closed
string coordinates which are consistent with fermionization of
those closed string bosonic degrees of freedom generally involve
such a projection, and the details of the projection depend on the
specific compactification.

 \subsection{Boundary States}

 The boundary state is a state of closed string theory which
 represents the interaction of closed strings with a D-brane. It is a
 closed string state which is annihilated by the boundary condition
 that would be imposed on an open string embedding function when the
 open string world-sheet ends on the D-brane world-volume. Depending
 on whether the open string embedding function is longitudinal or
 transverse to the brane, the boundary condition is Neumann or
 Dirichlet, respectively. Let us assume that the cylindrical
 world-sheet of the closed string encounters a D-brane when the
 world-sheet coordinate $\tau=0$.  The Neumann and Dirichlet boundary
 states for the bosonic string obey\footnote{The first of these is
 equivalent to usual Dirichlet condition
 $\partial_\tau\phi(0,\sigma)|{D}\rangle=0$ plus the additional condition that
 $(x_L-x_R)|{ D}\rangle=0$ which we are free to impose.}
 \begin{subequations}
 \label{generallabel}
 \begin{eqnarray}
 \phi_L(0,\sigma)\left|N \right\rangle&=&\phi_R(0,\sigma)\left|
 N \right\rangle, \label{LR:a}\\
 \phi_L(0,\sigma)\left| D \right\rangle&=&-\phi_R(0,\sigma)
 \left|D \right\rangle. \label{LR:b} \end{eqnarray}
 \end{subequations}

 The conditions (\ref{LR:a}) and (\ref{LR:b}) are solved by
 \begin{subequations}
 \label{generallabel}
 \begin{eqnarray}
 \left| N \right\rangle &=& \sum_{p_L} \prod_{n=1}^\infty \exp\left(
 -\frac{1}{n}\beta_{-n}\tilde\beta_{-n}\right)\left|
 p_L,-p_L\right\rangle,   \\
 \left| D\right\rangle &=& \sum_{p_L}\prod_{n=1}^\infty\exp \left(
 \frac{1}{n}\beta_{-n}\tilde\beta_{-n}\right)\left|p_L,p_L\right\rangle.
 \end{eqnarray}
 \end{subequations}
 In the fermion representation with Eqs.~(\ref{bosform:a}) and
 (\ref{bosform:b}), Eqs.~(\ref{LR:a}) and (\ref{LR:b}) imply that the
 boundary states satisfy
 \begin{subequations}
 \label{generallabel}
  \begin{eqnarray}
  \psi_L(0,\sigma)~\left|N \right\rangle &=&
  i\psi^{\dagger}_R(0,\sigma) \left|N \right\rangle~,~~
 \psi_L^{\dagger}(0,\sigma) \left|N \right\rangle=
  i \psi_R(0,\sigma)~\left| N\right\rangle, \\
  \psi_L(0,\sigma)~\left| D\right\rangle &=&
 - i\psi_R(0,\sigma) \left| D\right\rangle~,~~
 \psi_L^{\dagger}(0,\sigma) \left| D\right\rangle=
 - i \psi_R^{\dagger}(0,\sigma)~\left| D\right\rangle.
 \end{eqnarray}
 \end{subequations}
 Note that we have chosen cocycles in such a way
 that they cancel from these boundary
 equations.

 A solution of these equations  in the NS-NS sector is
 \begin{subequations}
 \label{generallabel}
 \beq
 \left| N\right\rangle_{NS} &=& \prod_{n=1/2}^\infty e^{i
 \left(\psi_{-n}^{\dagger}\tilde \psi^{\dagger}_{-n}+ \psi_{-n}\tilde
 \psi_{-n}\right)}\left|0\right\rangle_{NS}, \label{nns}\\
 \left| D\right\rangle_{NS} &=& \prod_{n=1/2}^\infty e^{-i
 \left(\psi_{-n}\tilde \psi^{\dagger}_{-n}+ \psi^{\dagger}_{-n}\tilde
 \psi_{-n}\right)}\left|0\right\rangle_{NS} \label{dns}
 \eeq
 \end{subequations}
 and in the RR-sector,
 \begin{subequations}
 \label{generallabel}
 \begin{eqnarray}
 \left| N\right\rangle_R &=& \prod_{n=0}^\infty e^{i
 \psi_{-n}^{\dagger}\tilde \psi^{\dagger}_{-n}} \cdot e^{i  \psi_{-n}\tilde
 \psi_{-n})} \left|--\right\rangle, \label{nr}
 \\
 \left| D\right\rangle_R &=& \prod_{n=0}^\infty e^{-i \psi_{-n}\tilde
 \psi^{\dagger}_{-n}} \cdot e^{-i \psi^{\dagger}_{-n}\tilde
 \psi_{-n}}\left|-+\right\rangle. \label{dr}
 \end{eqnarray}
 \end{subequations}

 \subsection{Open String Partition Functions}

 Now that we have constructed boundary states in the fermion representation, it
 is possible to compare open string partition functions. The boson Hamiltonians
 are given in (\ref{ham1}). In the boson theory, the partition function
 for the open string coordinate with Neumann boundary condition is
 \begin{equation}\label{bosneumann}
 Z_{NN}^B[q]=\left\langle N\right| q^{(L_0^B+\tilde L^B_0) }\left| N
 \right\rangle = \sum_{p_L} q^{p_L^2-\frac{1}{12}} \prod_{n=1}^\infty
 \frac{1}{1-q^{2n}}.
 \end{equation}
 The momentum $p_L$ here is summed over all integers and half-odd
 integers.  We see this from the following argument:
   We recall that, with the periodic identification (\ref{periodic}) of
   the boson at $R=1/\sqrt{2}$, the momenta are quantized so that they
   obey the constraints in Eq.~(\ref{momentumintegers}). In the Neuman
   state, $p_R=-p_L$ comes from the Neuman boundary condition.
   Therefore $p_L+p_R=0$ and the first condition, that
   $p_L+p_R$ be even, is automatic.  Further, the second condition,
   that $p_L-p_R$ must be an integer,   tells
   us that $2p_L=$integer. Thus, $p_L$ in the summation in (\ref{bosneumann})
   can take any integer or any half-odd integer value.  We will now
   show that the boson states where $p_L$ is an integer correspond to
   fermion NS-NS states and those where $p_L$ is half-odd-integer
   correspond to fermion
   R-R states.

 In the NS-NS sector the Hamiltonians are given in
 (\ref{ham2}). The NS-NS contribution to the  partition function is
 \begin{eqnarray} \label{nsnn}
 Z_{NN}^{NS}[q] &=&~~~_{NS}\left\langle N \right|q^{(L^{NS}_0+\tilde
 L^{NS}_0)} \left| N \right\rangle_{NS}
 =q^{-\frac{1}{12}}\prod_{n=1/2}^\infty
 \left( 1+q^{2n}\right)^2 \\
 &=& \sum_{m\in{\bf Z}} q^{m^2-\frac{1}{12}} \prod_{n=1}^\infty
 \frac{1}{1-q^{2n}}. \nn
 \end{eqnarray}
 In the last step, we used the Jacobi triple product identity
 (\ref{jacobi}) with $z=1$. Comparing the partition functions, we
 find that the Neumann state in the NS-NS sector contains only the
 bosonic states where $p_L$ is an integer.

To get the states where $p_L$ is a half-odd-integer, we must consider
the R-R sector.  The R-R sector Hamiltonians are given in (\ref{ham3}) and the  R-R
 contribution to the
 partition function is \beq Z^R_{NN}[q] &=& ~~_R \langle{N} \vert q^{
 \left(L^R_0+\tilde L^R_0\right)}
 \vert {N} \rangle_R \nonumber \\
 &=& 2 q^{\frac{1}{6}}\prod_{n=1}(1+q^{2 n})^2 = \sum_m q^{
 \left(\frac{2m +1}{2}\right)^2 -\frac{1}{12}} \prod_{n=1}^\infty
 \frac{1}{(1- q^{2 n})}. \eeq Indeed we see that, the R-R sector of the fermion
 theory corresponds to the boson states where $p_L$ is a
 half-odd-integer.

 Then, in order that the fermion theory produce the
 correct partition function, we need both the NS-NS and R-R sectors,
 \begin{equation}
  Z^B_{NN}[q] ~=~ Z^{NS}_{NN}[q]  ~+~Z^R_{NN}[q]
 \end{equation}
 Here, we could have completed the projection
 analogous to Eq.~(\ref{gso}) by
 including the projector $\frac{1}{2}\left( 1+(-1)^{p_L+p_R}\right)$ in the
 traces.  However,  in both the NS-NS and R-R
 fermion Neumann boundary states, the total $p_L+p_R$, which is the
 difference of right and left-handed fermion numbers, is zero.  Thus,
 the projection would have trivial effect and we have not done it explicitly.

 In summary, the description of the open string coordinate with Neumann
 boundary conditions in fermion variables requires use of both the
 NS-NS and the R-R sectors and the projection onto states with even total momentum,
 which is trivial in this case.

 The boson partition function
 for Dirichlet states is
 \beq Z^B_{DD}[q] &=& \langle D| q^{\left(L^B_0+\bar L^B_0\right)}|D\rangle \nn\\
&=& \sum_{p_L} q^{\left[ p_L^2 - \frac{1}{12}\right]} \prod_{n=1}
     \frac{1}{1-q^{2n}}. \eeq
     Now, since $p_L+p_R$ must be  an even integer and $p_L=p_R$,
     $p_L$ must be an integer.  There are no half-odd-integer momentum states
     in this case.

     By some simple algebra we find the partition
     function $Z_{DD}[q]$ in the fermion theory, using the fermion
     boundary states $|D\rangle$ Eq.~(\ref{dns}) and Eq.~(\ref{dr}). In
     the NS-NS sector
      \beq 
      Z^{NS}_{DD}[q] &=& \langle D|\frac{1}{2}\left(1+(-1)^{p_L+p_R}\right)
      q^{\left(L_0^{N}+\tilde L_0^{NS}\right)}|D\rangle\nonumber \\
      &=& q^{-\frac{1}{12}}
     \prod_{n=\half} (1+ q^{2n})^2 = \sum_{m \in {\bf Z}} q^{\left[
     m^2- \frac{1}{12}\right]} \prod_{n=1} \frac{1}{1-q^{2n}}. \eeq
  Note that $p_L+p_R$ is an even integer in all NS-NS Dirichlet states
  so the projection is a trivial operation here.
  However, because of fermion zero modes, it is odd in R Dirichlet
  states, so the R-R sector partition function must vanish,
       \beq Z^{R}_{DD}[q] =\langle D|\frac{1}{2}\left(1+(-1)^{p_L+p_R}\right)
       q^{\left(L_0^{R}+\tilde L_0^{R}\right)}|D\rangle =0. \eeq
 We see again that the projected combination of fermion NS-NS and R-R sectors
 indeed produces the boson partition function for open strings.

 \vskip .75cm \section{Rolling Tachyon and Fermionization}

After the Wick rotation, the rolling tachyon is described by the boundary
 conformal field theory (We choose $B=0$ and $A=\frac{g}{2}$.)
 \beq S = \frac{1}{4\pi} \int d\tau
 d\sigma \p_a X \p_a X -ig \pi \oint \frac{d\s}{2 \pi} e^{iX}.
 \label{action} \eeq
 Note that the factor of $i$ in front of the interaction comes
 from the fact that the world-sheet coordinates are also Wick
 rotated to Euclidean space and that the boundary integration variable $\sigma$
 is a spatial coordinate here, so is not Wick rotated.

 It is this theory which is solvable and which we
 shall represent using fermions.  We will do this by constructing a
 boundary state.  It is a state of closed string theory which is
 annihilated by the open string theory boundary condition,
 \begin{equation}
 \left.\left(
 \frac{1}{2\pi}\frac{\partial}{\partial\tau}X(\tau,\sigma)
 +\frac{g}{2} :e^{iX(\tau,\sigma)}:\right)\right|_{\tau=0}~\left|
 {B}\right\rangle=0. \label{bs}
 \end{equation}
 This is the boundary condition that one would impose in order to
 avoid surface contributions to the equation of motion when it is
 derived from varying the action (\ref{action}). Our goal is to find
 the boundary state $\left|{\cal B}\right\rangle$ which satisfies this
 equation.

 There is an important obstacle to introducing fermions in the most
 straightforward way. The periodic potential for the boson
 $X(\tau,\sigma)$ in (\ref{action}) has the wrong period to
 produce the fermionized
 closed string degrees of freedom which we discussed in the last
 Section. Recall that, in order to have a sensible fermion
 representation, the boson had to be identified with period
 $\phi(\tau,\sigma)\sim\phi(\tau,\sigma)+\sqrt{2}\pi$. Here the
 period is $X(\tau,\sigma)\sim X(\tau,\sigma)+2\pi$.

 Following ref.~\cite{pol}, we observe that this problem can be
 fixed by doubling the boson
 degrees of freedom.  Accordingly, we introduce a free boson
 $Y(\tau,\sigma)$ such that the action (\ref{action}) is replaced by

 \begin{equation} \label{action1}
 S=\frac{1}{4\pi}\int d\tau d\sigma \left(\partial_a X\partial_a X
 +\partial_a Y\partial_a Y\right) - ig \pi \int \frac{d\sigma}{2\pi}
 e^{iX(0,\sigma)}.
 \end{equation}
 We will assume that the Y-boson has a Dirichlet boundary condition
 and that, like $X$, it is periodically identified with
 $$ Y\sim Y+2\pi~~.$$ We will then construct a double boundary state,
$\left| {B,D}\right\rangle$ which obeys the boundary condition for $X$ as
in Eq.(\ref{bs}) as well as an additional condition for $Y$,
\begin{subequations} \label{generallabel}
\begin{eqnarray}\label{bs1} \left(
\frac{1}{2\pi}\frac{\partial}{\partial\tau}X(0,\sigma)
+\frac{g}{2}:e^{iX(0,\sigma)}:\right)~\left| {B,D}\right\rangle &=& 0, \\
Y(0,\sigma)~\left|B,D\right\rangle &=& 0. \label{bs2} \end{eqnarray}
\end{subequations} Since $X$ and $Y$ do not interact, it will be
always be easy to separate the contribution of the $Y$-field from
matrix elements.  Its contribution factors into one for a free boson
with a Dirichlet boundary condition which is easily identified and
discarded in the amplitudes that we compute.

The Dirichlet boundary condition for $Y$ in (\ref{bs2}) implies
 \begin{equation}
 \left[
 \partial_\tau Y_L(0,\sigma) -\partial_\tau
 Y_R(0,\sigma)
 \right]
 \left|{B,D}\right\rangle=0.
 \label{diri}\end{equation}
 Using (\ref{bs2}) and (\ref{diri}), we can re-write (\ref{bs1}) as
 \begin{eqnarray} \label{phi1}
 \left[ \frac{1}{2\pi}\frac{\partial}{\partial\tau}
 \left(X_L(0,\sigma)+Y_L(0,\sigma)\right)
       +\frac{1}{2\pi}\frac{\partial}{\partial\tau}
 \left(X_R(0,\sigma)-Y_R(0,\sigma)\right)+~~~~~~~~~~~~~
  \right. \nonumber \\ \left.
 +\frac{g}{2}:e^{i ( X_L(0,\sigma)+Y_L(0,\sigma) ) }: ~:e^{ i(
 X_R(0,\sigma)+Y_R(0,\sigma) ) } :  \right]~\left| {B,D}\right\rangle = 0.
 \label{phi2}\end{eqnarray}
 An infinite factor from normal ordering
 $e^{iY(0,\sigma)}=e^{iY_L(0,\sigma)}e^{iY_R(0,\sigma)}$ has been
 absorbed into $g$.  We can re-express the boson fields as the
 canonically normalized pair
 \begin{subequations}
 \label{generallable}
 \begin{eqnarray}\label{phi1}
 \phi_{1L}(\tau,\sigma)=\frac{X_L(\tau,\sigma)+Y_L(\tau,\sigma)}{\sqrt{2}}
 ~~,~~
 \phi_{1R}(\tau,\sigma)=\frac{X_R(\tau,\sigma)+Y_R(\tau,\sigma)}{\sqrt{2}}
 \\ \label{phi2}
 \phi_{2L}(\tau,\sigma)=\frac{X_L(\tau,\sigma)-Y_L(\tau,\sigma)}{\sqrt{2}}
 ~~,~~
 \phi_{2R}(\tau,\sigma)=\frac{X_R(\tau,\sigma)-Y_R(\tau,\sigma)}{\sqrt{2}}.
 \end{eqnarray}
 \end{subequations}
  In terms of these, we obtain
 \begin{subequations}
 \label{generallabel}
 \begin{eqnarray}
 \left({\partial_\tau}\phi_{1L}(0,\sigma)
 +{\partial_\tau}\phi_{2R}(0,\sigma)
 +\frac{g\pi}{\sqrt{2}}:e^{\sqrt{2}i\phi_{1L}(0,\sigma)}:~:e^{\sqrt{2}i
 \phi_{1R}(0,\sigma)}:\right)~\left| {B,D}\right\rangle=0, \label{inter1}
 \\ \label{inter2}
\left[ \phi_{1L}(0,\sigma)+\phi_{1R}(0,\sigma)-
\phi_{2L}(0,\sigma)-\phi_{2R}(0,\sigma)\right]~\left|
 {B,D}\right\rangle=0.
 \end{eqnarray}
 \end{subequations}
 Note that, normal ordering for $(X,Y)$ implies normal ordering for
 $(\phi_1,\phi_2)$.

 The problem that has been solved here is the following: Recalling our
 discussion of fermionizing a compact boson surrounding
 Eqs.~(\ref{compact1}) and (\ref{compact2}), we see that, when written in terms of
 the variables $\phi_1$ and $\phi_2$, the equation
 for the boundary state (\ref{inter1}), has the correct periodicity for identification
 of the exponentials with fermion operators.  This is in contrast with
 its previous form (\ref{bs}) in terms of $X$  .The
 definitions (\ref{phi1}) and (\ref{phi2}) of $\phi_1$ and $\phi_2$,
 imply the identifications $\phi_1\sim
 \phi_1+\sqrt{2}\pi$ and $\phi_2\sim\phi_2+\sqrt{2}\pi$ when
 $X\sim X+2\pi$ and $Y\sim Y+2\pi$.

  \subsection{Fermion representation of the doubled bosons}

  We will now need to generalize fermionization to
 the doubled system. There are two species of left and right-handed
 fermions, together with their conjugates,
 \begin{subequations}
 \label{generallabel}
 \begin{eqnarray}
 \psi_{1L}(z)~&=&~\eta_{1L}~:e^{-i\sqrt{2}\phi_{1L}(z)}:~,
 \psi_{1R}(\barz)~=~\eta_{1R}~:e^{i\sqrt{2}\phi_{1R}(\bar z)}:,\\
 \psi_{2L}(z)~&=&~\eta_{2L}~:e^{i\sqrt{2}\phi_{2L}(z)}:~,
 \psi_{2R}(\barz)~=~\eta_{2R}~:e^{-i\sqrt{2}\phi_{2R}(\bar
 z)}:.
 \end{eqnarray}
 \end{subequations}
 The explicit expression of the cocycles is given in the Appendix.
 The cocycles have been carefully adjusted so that all independent
 fields anti-commute.  Some further freedom is used to adjust them so that they do
 not appear explicitly in other formulae, such as the fermion boundary conditions
 that we will find below.  The fermion currents are given by
 \begin{subequations}
 \label{generallabel}
 \begin{eqnarray}\label{currents1}
 i\sqrt{2}\partial_\tau
 \phi_{1L}(\tau,\sigma) &=&:\psi_{1L}^{\dagger}(\tau,\sigma)\psi_{1L}(\tau,\sigma):
 ~,~ -i\sqrt{2}\partial_\tau
 \phi_{1R}(\tau,\sigma)=:\psi_{1R}^{\dagger}(\tau,\sigma)\psi_{1R}(\tau,\sigma):\\
 \label{currents2}
 -i\sqrt{2}\partial_\tau
 \phi_{2L}(\tau,\sigma) &=& :\psi_{2L}^{\dagger}(\tau,\sigma)\psi_{2L}(\tau,\sigma):~,~
 i\sqrt{2}\partial_\tau
 \phi_{2R}(\tau,\sigma)=:\psi_{2R}^{\dagger}(\tau,\sigma)\psi_{2R}(\tau,\sigma):.
 \end{eqnarray}
 \end{subequations}

 In order to construct the rolling tachyon boundary state we shall
 need a fermion representation of $|{ N,D}\rangle$. In boson variables it
obeys the boundary condition
 \begin{equation}\label{doublebs}  \left(X_L(0,\sigma)-X_R(0,\sigma)\right)
 \left| N,D\right\rangle=0 ~~,~~
 \left(Y_L(0,\sigma)+Y_R(0,\sigma)\right)\left| N,D\right\rangle=0 \end{equation}
 or
\begin{equation} \label{119}\phi_{1L}(0,\sigma)\left|{
 N,D}\right\rangle=\phi_{2R}(0,\sigma) \left|{ N,D}\right\rangle=0 ~~,~~
 \phi_{2L}(0,\sigma)\left|{ N,D}\right\rangle=\phi_{1R}(0,\sigma) \left|{
 N,D}\right\rangle
 \end{equation}
In fermion variables, 
\begin{subequations}
\label{generallabel}
\beq \label{121} \psi_{1L}(0,\sigma)|{
 N,D}\rangle &=& \psi_{2R} (0,\sigma) |{ N,D}\rangle,\quad
 \psi_{2L}(0,\sigma)|{ N,D}\rangle = - \psi_{1R}(0,\sigma)|{ N,D}\rangle, \\
\label{122}
 \psi^{\dagger}_{1L}(0,\sigma)|{ N,D}\rangle &=& -
 \psi^{\dagger}_{2R}(0,\sigma)|{ N,D}\rangle,\quad
 \psi^{\dagger}_{2L}(0,\sigma)|{ N,D}\rangle =
 \psi^{\dagger}_{1R}(0,\sigma)|{ N,D}\rangle . \eeq
 \end{subequations} 
 We will construct this state (and others) explicitly in the next Section.  
 Note that Eq.~(\ref{119}) already implies the boundary condition in Eq.~(\ref{inter2}).

 Now, Eq.~(\ref{inter1}) becomes
 \begin{eqnarray}
 \left[:\psi_{1L}^{\dagger}(0,\sigma)\psi_{1L}(0,\sigma): +
 :\psi_{2R}^{\dagger}(0,\sigma)\psi_{2R}(0,\sigma):+i g \pi
 :\psi_{1L}^{\dagger} (0,\sigma)\psi_{1R} (0,\sigma):\right]\left| {
 B,D}\right\rangle=0. \label{inter3}
 \end{eqnarray}
 A solution of this equation is
 \begin{equation}
 \left| { B,D}\right\rangle~=~\exp\left(i g\pi \int \frac{d\s}{2\pi}
 :\psi_{1L}^{\dagger}(0,\sigma)\psi_{2L}(0,\sigma):\right)~\left|{
 N,D}\right\rangle. \label{solution}
 \end{equation} 
 To see this, note that, substituting (\ref{solution}) into (\ref{inter3}) implies
 \begin{eqnarray}
 \left[:\psi_{1L}^{\dagger}(0,\sigma)\psi_{1L}(0,\sigma): +
 :\psi_{2R}^{\dagger}(0,\sigma)\psi_{2R}(0,\sigma):-
 ~~~~~~~~~~~~~~~~~~~~~~~~~~~~~~~~~~~~~~~~~~~~~~ \right.\nonumber
 \\ \left.  +i g\pi \psi_{1L}^{\dagger}(0,\sigma) \left(\psi_{1R}(0,\sigma)
 +\psi_{2L}(0,\sigma)\right) \right]~\left| { N,D}\right\rangle=0.
 \label{inter4}
 \end{eqnarray}
 Indeed, using Eqs.~(\ref{121}) and (\ref{122}) we see that
 this is an identity as both the combination of the first two terms and the last
 term annihilate the state   $\left|{ N,D}\right\rangle$.

 In the next
 Section, we will give an explicit construction of the Neumann-Dirichlet
 state $|N,D\rangle$.   Then, combining the result with (\ref{solution})
 will finally yield the complete solution of the problem of finding a
 fermion representation of the rolling tachyon boundary state.

 A further observation is that the fermion boundary condition for
 the rolling tachyon boundary state can be presented as
 \begin{subequations}
 \label{generallabel}
 \beq \label{1211} \left(\psi_{1L}(0,\sigma)-i\pi
 g\psi_{2L}(0,\sigma)\right)|{
 B,D}\rangle &=& \psi_{2R} (0,\sigma) |{ B,D}\rangle,
 \\
 \quad
 - \psi_{2L}(0,\sigma)|{ B,D}\rangle &=& \psi_{1R}(0,\sigma)|{ B,D}\rangle, \\
 \label{1221}
 \left(\psi^{\dagger}_{2L}(0,\sigma)+i\pi g\psi^{\dagger}_{1L}(0,\sigma)
 \right)|{ B,D}\rangle &=&
 \psi^{\dagger}_{1R}(0,\sigma)|{ B,D}\rangle,
 \\
 \quad
 -\psi^{\dagger}_{1L}(0,\sigma)|{ B,D}\rangle &=& 
 \psi^{\dagger}_{2R}(0,\sigma)|{ B,D}\rangle
 . \eeq
 \end{subequations}

 \subsection{Quantum theory of the bosons $(\phi_1,\phi_2)$}

 To fix our conventions, we shall examine the free boson fields
 $(\phi_{1},\phi_2)=\left(\frac{1}{\sqrt{2}}(X +
 Y),\frac{1}{\sqrt{2}}(X-Y)\right)$
 with action
 \beq
 S=\frac{1}{4\pi}\int d\tau d\sigma \left(
 (\partial\phi_1)^2+(\partial\phi_2)^2\right) = \frac{1}{4\pi}\int d\tau
 d\sigma \left((\partial X)^2+(\partial Y)^2\right).
 \eeq
 They
 have the mode expansion
 \begin{subequations}
 \label{generallabel} \beq \phi_{aL}(\t+i\s) &=& \frac{1}{\sqrt{2}}
 \chi_{aL} - \frac{i}{\sqrt{2}}\pi_{aL} (\t+i\s)+
 \frac{i}{\sqrt{2}}\sum_{n\not= 0}
 \frac{1}{n} \b_{an} e^{-n(\t+i\s)}, \\
 \phi_{aR}(\t-i\s) &=& \frac{1}{\sqrt{2}} \chi_{aR} -
 \frac{i}{\sqrt{2}} \pi_{aR} (\t-i\s)+ \frac{i}{\sqrt{2}}\sum_{n\not=
 0} \frac{1}{n} \tilde{\b}_{an} e^{-n(\t-i\s)}. \eeq
 \end{subequations}
 The non-vanishing commutation relations are \beq
 \left[\chi_{aL}, \pi_{bL} \right] &=& i\delta_{ab},
 \quad \left[\chi_{aR}, \pi_{bR} \right] = i\delta_{ab}, \\
 \left[\b_{an}, \b_{bm} \right] &=& n \delta(n+m)\delta_{ab}, \quad
 \left[\tilde{\b}_{an}, \tilde{\b}_{bm} \right] =n
 \delta(n+m)\delta_{ab}. \nn \eeq

The doubled fermions have mode expansions
\begin{subequations}
\label{generallabel}
\begin{eqnarray}
\psi_{aL}(\tau+i\sigma)=\sum_n \psi_{an} e^{-n(\tau+i\sigma)}
~~,~~\psi^{\dagger}_{aL}(\tau+i\sigma)=\sum_n \psi_{an}^{\dagger}
e^{-n(\tau+i\sigma)} \\
\psi_{aR}(\tau-i\sigma) = \sum_n \tilde\psi_{an} e^{-n(\tau-i\sigma)}
~~,~~
\psi_{aR}^{\dagger}(\tau-i\sigma) = \sum_n \tilde\psi_{an}^{\dagger} e^{-n(\tau-i\sigma)}
\end{eqnarray}
\end{subequations}
where the non-vanishing anti-commutators are
\begin{equation}
\left\{ \psi_{am},\psi^{\dagger}_{bn}\right\}=\delta_{ab}\delta_{m+n}
~~,~~\left\{ \tilde\psi_{am},\tilde\psi^{\dagger}_{bn}\right\}=\delta_{ab}\delta_{m+n}
    \end{equation}
  In our discussion of a fermion representation of a single boson,
  we noted that the momenta $(\pi_{1L},\pi_{1R})$
  are identified with fermion numbers, and they also control the
  periodicity of the fermions under $\sigma\to\sigma+2\pi$.
  This quantization of the fermion number and therefore
  the momenta requires a periodic
  identification of the boson.  For example,    if
  $\phi_{1L}\sim\phi_{1L}+\sqrt{2}\pi$, then $p_{1L}$ is quantized
either as integers or half-odd-integers.
  Integer momentum corresponds an NS fermion whereas
  half-odd-integer momentum corresponds to an $R$ fermion.
  Then, for example, to get all of the states of the compact boson
  $\phi_1=\phi_{1L}+\phi_{1R}$ with $\phi_1\sim\phi_1+\sqrt{2}\pi$
  we must take the NS-NS and R-R states of the fermions $(\psi_{1L},\psi_{1R})$
  and further project onto states where
  $\pi_1=\pi_{1L}+\pi_{1R}$  is an even integer.

A similar construction obtains all of the states of the
  compact boson $\phi_2$ and the total space of all states is the direct
  product of these representations. However, these are not quite the correct states 
  to describe the theory which has compact bosons $X$ and $Y$
  with identifications $X\sim X+2\pi$ and $Y\sim Y+2\pi$.
To  correctly describe them, we need the states where the momenta
of $X$ and $Y$, rather than $\phi_1$ and
$\phi_2$ are quantized appropriately. It is easy to see that these
are the states where
\begin{subequations}
\label{generallabel}
\begin{eqnarray}\label{221}
\pi_{1L}+\pi_{2L}+\pi_{1R}+\pi_{2R}=~2m_X
\\ \label{222}
\pi_{1L}-\pi_{2L}+\pi_{1R}-\pi_{2R}=~2m_Y
\\ \label{223}
 \pi_{1L}+\pi_{2L}-\pi_{1R}-\pi_{2R}=~2w_X
\\ \label{224}
 \pi_{1L}-\pi_{2L}-\pi_{1R}+\pi_{2R}=~2w_Y
\end{eqnarray}
\end{subequations}
where $(m_X,m_Y,w_X,w_Y)$ are
 integers.
 In order to satisfy Eqs.~(\ref{221})-(\ref{224}), there are two
 possibilities:   all momenta
 $(\pi_{1L},\pi_{1R},\pi_{2L},\pi_{2R})$ must be integers, or
all momenta $(\pi_{1L},\pi_{1R},\pi_{2L},\pi_{2R})$
 must be half-odd integers.  As a result, both fermions
 $(\psi_{1L},\psi_{1R})$ and $(\psi_{2L},\psi_{2R})$ must
 be   NS-NS,
 or both must be  R-R.  Then, in addition to this constraint,
 to obtain all states satisfying satisfy Eqs.~(\ref{221})-(\ref{224}),
 it is sufficient to project onto states where one of the combinations in
 (\ref{221})-(\ref{224}) is even, for example states where
 $\pi_{1L}-\pi_{2L}-\pi_{1R}+\pi_{2R}$ is even.  This
 can be done with the projection operator
  \beq   P=
  \frac{1}{2}\left[  1+(-1)^{\pi_{1L}-\pi_{1R}
   -\pi_{2L}+\pi_{2R}}\right]
  \eeq
  These are the states where, in our convention for identifying
  momenta and fermion numbers, the total fermion
  number is even.

 To confirm that this is the correct projection of the fermion
 states, let us compare partition functions.
 The partition function for the $X$-boson is
 \beq
 Z[q,\bar q] &=&
  {\rm Tr}\left[ q^{L_{X0} } \bar q^{\tilde L_{X0} }  \right]  
  \nonumber \\
  &=& \sum_{p_{XL} ,p_{XR} }
 q^{\frac{1}{2}p_{XL}^2  }
 \bar q^{\frac{1}{2}  p_{XR}^2 }  \cdot
 \left| q^{-\frac{1}{24}} \prod_1^\infty \frac{1}{1-q^n}
 \right|^2 \\
 &=& \left[
 \sum_{m,n\in{\cal Z}} q^{\frac{1}{4}(m+n)^2}\bar
 q^{\frac{1}{4}(m-n)^2
 }\right]\cdot\left|q^{-\frac{1}{24}}  \prod_1^\infty \frac{1}{1-q^n}
 \right|^2 \nonumber
 \eeq
Since both $m$ and $n$ are integers, $m+n$ and $m-n$ must either both be even or both odd.
Thus,
\beq
 Z[q,\bar q] =  \left[
  \left|\sum_{n\in{\cal Z}} q^{n^2}\right|^2
   +\left|\sum_{n\in{\cal Z}}  q^{(n+\frac{1}{2})^2}\right|^2
  \right]\cdot\left|q^{-\frac{1}{24}}  \prod_1^\infty \frac{1}{1-q^n}
  \right|^2
\eeq
 and the partition function for the $(X,Y)$ system is just the
 square, $\left( Z[q,\bar q]\right)^2$.

We will now see if we recover this result in the fermion theory. In
the projected NS-NS sector we get
\beq
Z_{NS}[q,\bar q]=
 {\rm Tr}\left[ P q^{L^{NS}_{0}}\bar q^{\tilde L^{NS}_{0}}\right]=
\frac{1}{2}
\left|  q^{-\frac{1}{24}}\prod_0^\infty(1+q^{n+\frac{1}{2}})\right|^4+
\frac{1}{2}\left|q^{-\frac{1}{24}}\prod_0^\infty(1-q^{n+\frac{1}{2}})
\right|^4.
\eeq
Here, the first term comes from the $1$ in the projection operator.
The second term comes from the second term in the projection
operator, which changes the sign of the excites states. The Jacobi
identity leads to
\beq
Z_{NS}[q,\bar q]=    \frac{1}{2} \left(\left|\sum_{n\in{\cal
Z}}q^{\frac{1}{2}n^2}\right|^4+ \left|\sum_{n\in{\cal
Z}}(-1)^nq^{\frac{1}{2}n^2}\right|^4\right) \left|
q^{-\frac{1}{24}}\prod_{n=1}^\infty \frac{1}{1-q^n}\right|^4
\eeq
and in the projected R-R sector,
\beq
 Z_{R}[q,\bar q] &=&
  {\rm Tr}\left[ P q^{L^R_{0}}\bar q^{\tilde L^R_{0}}\right]=
 \frac{1}{2}
 \left|  q^{\frac{1}{12}}2 \prod_1^\infty(1+q^{n})\right|^4 \nonumber \\
&=&  \frac{1}{2}
     \left|\sum_{n\in{\cal Z}}q^{\frac{1}{2}\left(n+\frac{1}{2}\right)^2}\right|^4
    \left| q^{-\frac{1}{24}}\prod_{n=1}^\infty 
    \frac{1}{1-q^n}\right|^4.
\eeq
Here, we have used the
    identities in Eqs.~(\ref{id1})-(\ref{id3}).
Indeed, by comparing the partition functions,
\beq
(Z[q,\bar q])^2 &=&  Z_{NS}[q,\bar q]+Z_R[q,\bar q] \nonumber \\
&=& \left(
1+8|q|^{\frac{1}{2}}+16|q|+4q+4\bar q+16q^{\frac{1}{4}}\bar q^{\frac{5}{4}}+
16q^{\frac{5}{4}}\bar q^{\frac{1}{4}}+4q^2+4\bar q^2+16q\bar q+...\right)
\left| q^{-\frac{1}{24}}\prod_{n=1}^\infty 
\frac{1}{1-q^n}\right|^4 \nonumber \\
&=&\left( 1+4|q|^{\frac{1}{2}}+2q+2\bar
                 q+\ldots\right)^2
\left| q^{-\frac{1}{24}}\prod_{n=1}^\infty 
\frac{1}{1-q^n}\right|^4 .
\eeq

 \subsection{Current Algebra}

 It is well known that the vertex operator $:e^{2iX_L(z)}:$, together
 with  $:e^{-2iX_L(z)}:$ and $i\p_\tau X(z)$ form a level-one $SU(2)$ Kac-Moody
  algebra when  the theory is defined on a circle with the
 self-dual radius $R_X=1$ \beq J^+(z) &=& :e^{2iX_L(z)}:,\quad J^-(z) =
 :e^{-2iX_L(z)}:, \quad
 J^3(z) = iz\frac{\p X_L}{\p z}(z), \nn\\
 J^a(z)&=&\sum_n J^a_n \frac{1}{z^n} \nn\\
 \left[J^3_n, J^3_m \right] &=& \frac{n}{2} \delta_{n+m}, \quad
 \left[J^+_n, J^-_m \right] = 2J^3_{n+m} + n\delta_{n+m}, \\
 \left[J^3_n, J^+_m \right] &=& J^+_{n+m}, \quad \left[J^3_n, J^-_m
 \right] = -J^-_{n+m}. \nn \eeq This $SU(2)$ algebra plays an
 important role when one constructs the boundary states for the
 conformal field theory with the periodic boundary interaction. It is
 known \cite{call93,call94} that the boundary state can be written in
 terms of the Ishibashi states as \beq |{ B} \rangle = \sum_j
 \sum_{m\ge 0} \left(\begin{array}{c} j+m
 \\ 2m
 \end{array}\right)(i\pi g)^{2m} |j, m, m\rangle\rangle
 \eeq where $|j, m, m\rangle\rangle$ are the $SU(2)$ Ishibashi states
 and $j=0, \half, 1, \frac{3}{2}, \dots$.

 However, one drawback of this approach is that it is quite laborious
 to trace the explicit time dependence of the Ishibashi states beyond
 the first few levels. Furthermore,  it is almost impossible to work
 out the explicit time dependence of all Ishibashi states.  Because
 of this difficulty, a complete description of the long time behavior
 of the full boundary state is still lacking. (See, however, recent
 discussion on this point by Sen \cite{sen04}.)

 For practical purposes, it would often be more convenient to rewrite
 the boundary state in the closed string oscillator basis. In
 particular, it would be useful for the  study of its physical
 properties such as time evolution and closed string emission.
 However, it is quite difficult to express the Ishibashi states in
 general in such a way. An exception is the scalar sector which does
 not involve any oscillators. If we expand the boundary state in the
 bosonic oscillator basis, we obtain\beq |{ B} \rangle &=& f(t) |0,t
 \rangle + g(t)
 \a_{-1}\tilde{\a}_{-1}|0,t\rangle + \cdots, \\
 f(t)  &=& \sum_j (i\pi g )^{2j}\langle 0,t |j;j,j\rangle. \nn \eeq
 Adopting the phase convention for $|j;j,j\rangle$ as in
 \cite{sen02r}, we have \beq \label{scalarf} f(t)  = \sum_j (-\pi g
 e^t)^{2j}  = \frac{1}{1+ \pi g e^t } \eeq where we have done the
 reverse Wick rotation back to Lorentzian time.

 In the fermion theory, the degrees of freedom are doubled.  Indeed, there are two
 species of left-moving fermions $\psi_{1L}$ and $\psi_{2L}$ which we expect to
 form a doublet under $SU(2)_L$.  This $SU(2)_L$  Kac-Moody algebra is realized
 as the current algebra of the fermion currents
 \beq J^+_L(z) &=& :\psi^{\dagger}_{1L} (z)\psi_{2L}(z): =
 :\Psi^{\dagger}_L(z)\frac{(\s_1 +i\s_2)}{2} \Psi_L(z):, \nn\\
 J^-_L (z)&=& :\psi^{\dagger}_{2L}(z) \psi_{1L}(z):
 = :\Psi^{\dagger}_L(z)\frac{(\s_1-i\s_2)}{2} \Psi_L(z): \label{kacmoody1}\\
 J^3_L(z) &=& \half\left(:\psi^{\dagger}_{1L} (z)\psi_{1L}(z):- :\psi^{\dagger}_{2L}(z)
 \psi_{2L}(z):\right)=:\Psi^{\dagger}_L (z)\frac{\s_3}{2} \Psi_L(z): \nn \eeq where
 $$\Psi_L(z) = \left(\begin{matrix}\psi_{1L}(z)\cr
 \psi_{2L}(z)\cr\end{matrix}\right)  ~~,~~
 \Psi_R(\bar z) = \left(\begin{matrix}\psi_{1R}(\bar z)\cr
 \psi_{2R}(\bar z)\cr\end{matrix}\right)$$
 We might expect that the doubling of degrees of freedom comes with an
enhanced symmetry.
Indeed, if we decompose the
 complex fermions into real and imaginary parts, we find an
 $SO(4)_L=SU(2)_L\times SU(2)'_L$ algebra.   In terms of the complex
 generators, the $SU(2)_L$ algebra is (\ref{kacmoody1}) and $SU(2)'_L$ is
  \beq {J'}^+_L (z)&=& :\psi^{\dagger}_{1L} (z)\psi^{\dagger}_{2L}(z): =
  i:\Psi^{\dagger}_L(z) \frac{\sigma^2}{2} \Psi^{\dagger t}_L(z):, \label{kacmoody2}\nn\\
  {J'}^-_L (z)&=& :\psi_{2L}(z)  \psi_{1L}(z):
  = -i:\Psi^t_L(z)\frac{\sigma^2}{2} \Psi_L(z):  \\
  {J'}^3_L(z) &=& \half\left(:\psi^{\dagger}_{1L}(z) \psi_{1L}(z):+ :\psi^{\dagger}_{2L}(z)
  \psi_{2L}(z):\right)=\half:\Psi^{\dagger}_L(z)   \Psi_L(z): \nn \eeq
  The algebras (\ref{kacmoody1}) and (\ref{kacmoody2}) commute with
  each other and each is a level-1 $SU(2)$ Kac-Moody algebra.

Note that, because $\psi_{1L}$ and $\psi_{2L}$ have the same
boundary condition in the NS-NS and R-R sectors, both sets of
charges $J_0^a$ and ${J'}_0^a$ commute with the Hamiltonian and the
symmetries are not broken by boundary conditions. A similar
discussion applies to right-handed fermions.

 \vskip .75cm \section{Exact Boundary State for the Unstable D-Brane}

 In the Section II we studied the fermion representation of a
 conformal field theory with a single boson.  The Neumann and
 Dirichlet boundary conditions in the boson theory were shown to be
 realized as linear boundary conditions in the fermion theory and the
 boundary states were constructed explicitly.  In Section III we
 constructed the boundary state of the rolling tachyon as a certain
 global chiral SU(2) rotation of a mixed Neumann-Dirichlet state of
 the $X$ and $Y$ bosons.  It remains to complete our task of finding
 the rolling tachyon boundary state by finding the explicit
 Neumann-Dirichlet boundary state in the fermion theory.  We will
 address this problem in this Section.  The result will be a fermionic
 representation of the exact boundary state for the unstable D-brane.
 To begin, in the next Subsection, we will construct some simple
 boundary states with combinations of Neumann and Dirichlet boundary conditions for
 $X$ and $Y$.

 \vskip .75cm \begin{subsection}
 {\bf Some Simple Boundary States}
 \end{subsection}

 In this Subsection, we will consider mixed boundary states $|{ N}, {
 N}\rangle$, $|{ N}, { D}\rangle$, $|{ D}, { N}\rangle$ and $|{ D}, {
 D}\rangle$ where $N$ and $D$ denote Neumann and Dirichlet boundary
 conditions and the first and second label is the condition for the X
 and Y bosons, respectively.  We will translate the boundary conditions to
 conditions on the fermion variables.  We then use these conditions
 to construct the boundary states in the fermion representation.  In order to
 get all of the states of the boson theory, this must be done in both the
 NS-NS and R-R sectors of the fermion theory.  We will also show that these
 boundary states are related to each other by simple
 global $SU(2)_L\otimes SU(2)_L'$ rotations.

We begin with the $|N,N\rangle$ state which obeys the boundary conditions
\beq X_L (0,\sigma) |{ N},
 { N} \rangle = X_R(0,\sigma) |{ N},{ N}\rangle, \quad Y_L
 (0,\sigma)|{ N},{ N} \rangle = Y_R (0,\sigma) |{ N},{ N}\rangle,
\eeq
 This implies the fermion boundary conditions (see Appendix) \beq
 (\psi_{aL}(0,\sigma)- i\psi^{\dagger}_{aR}(0,\sigma)) |{ N}, {
 N}\rangle =0, \quad (\psi^{\dagger}_{aL}(0,\sigma)-
 i\psi_{aR}(0,\sigma)) |{ N}, { N}\rangle =0, \quad a = 1,2.
 \label{nnbc}\eeq
The boundary state which obeys these conditions is, in the NS-NS
sector,

\begin{equation}
|N,N\rangle_{NS}=\prod_{r=\half}^\infty
\left(1+i\psi^{\dagger}_{1,-r}\tilde\psi^{\dagger}_{1,-r}\right)
\left(1+i\psi_{1,-r}\tilde\psi_{1,-r}\right)
\left(1+i\psi^{\dagger}_{2,-r}\tilde\psi^{\dagger}_{2,-r}\right)
\left(1+i\psi_{2,-r}\tilde\psi_{2,-r}\right)~|0\rangle
\end{equation}
and in the R-R sector
\begin{equation}
|N,N\rangle_{R}=\prod_{n=0}^\infty
\left(1+i\psi^{\dagger}_{1,-n}\tilde\psi^{\dagger}_{1,-n}\right)
\left(1+i\psi_{1,-n}\tilde\psi_{1,-n}\right)
\left(1+i\psi^{\dagger}_{2,-n}\tilde\psi^{\dagger}_{2,-n}\right)
\left(1+i\psi_{2,-n}\tilde\psi_{2,-n}\right)|----\rangle.
\end{equation}
Here, we use the notation $|----\rangle=|--\rangle|--\rangle$ where the ordering of
zero mode state labels is
$\psi_{1L},\psi_{1R},\psi_{2L},\psi_{2R}$. Note that the
oscillator parts of the boundary states in the two sectors are
practically identical. A difference lies in the vacuum states.  In
the NS-NS sector we use the direct product of the NS-NS vacua for
each of the two species of fermions.  In the R-R sector, the
vacuum is degenerate and we must choose the combination of states
which obeys the boundary condition.

 Note that the boundary conditions only determine the boundary states
 up to a phase.  In the NS-NS sector, the phase has been fixed by requiring
 that $ \langle0|N,N\rangle_{NS}=1$, which matches the phase convention which we
 choose in the boson theory.  We will do this consistently
 in the NS-NS sector for all of the boundary states that we construct.
 In the RR-sector we shall discuss a similar matching of the phases in a later Section.
 That matching of phases is anticipated in the formulae for R sector
 states presented in this Section.

The boundary state $|{ N}, {
 D}\rangle$ has the boundary condition \beq X_L
 (0,\sigma) |{ N}, { D}\rangle = X_R (0,\sigma) |{ N}, { D}\rangle,
 \quad Y_L (0,\sigma) |{ N}, { D}\rangle =-Y_R (0,\sigma) |{ N}, {
 D}\rangle. \eeq In terms of fermions (see Appendix), \beq
 (\psi_{1L}(0,\sigma) - \psi_{2R}(0,\sigma) ) |N, D\rangle &=& 0,
 \quad (\psi_{2L}(0,\sigma) +\psi_{1R}(0,\sigma) ) |N, D\rangle =0,
 \label{fermionnd}\\ (\psi^{\dagger}_{1L}(0,\sigma) +
 \psi^{\dagger}_{2R}(0,\sigma) ) |N, D\rangle &=& 0, \quad
 (\psi^{\dagger}_{2L}(0,\sigma) - \psi^{\dagger}_{1R}(0,\sigma) ) |N,
 D\rangle =0. \nn \eeq

 The boundary state $|N, D\rangle$ is

\begin{equation}\label{bsnd1}
|N,D\rangle_{NS}=\prod_{r=\half}^\infty
\left(1+\tilde\psi_{1,-r}\psi^{\dagger}_{2,-r}\right)
\left(1-\tilde\psi^{\dagger}_{1,-r}\psi_{2,-r}\right)
\left(1+\psi^{\dagger}_{1,-r}\tilde\psi_{2,-r}\right)
\left(1-\psi_{1,-r}\tilde\psi^{\dagger}_{2,-r}\right) |0\rangle
\end{equation}
\begin{equation}\label{bsnd2}
|N,D\rangle_{R}=-i\prod_{n=0}^\infty
\left(1+\tilde\psi_{1,-n}\psi^{\dagger}_{2,-n}\right)
\left(1-\tilde\psi^{\dagger}_{1,-n}\psi_{2,-n}\right)
\left(1+\psi^{\dagger}_{1,-n}\tilde\psi_{2,-n}\right)
\left(1-\psi_{1,-n}\tilde\psi^{\dagger}_{2,-n}\right) 
|-+-+\rangle.
\end{equation}

The boson boundary condition for $|D,N\rangle$ is given by
\beq X_L (0,\sigma)|D,N\rangle = -X_R(0,\sigma) |D,N\rangle \quad
Y_L(0,\sigma) |D,N\rangle = Y_R(0,\sigma) |D,N\rangle, \eeq which
can be transcribed into the fermion boundary condition as \beq
\psi_{1L}(0,\sigma) |D, N\rangle &=&
\psi^{\dagger}_{2R}(0,\sigma)|D,N\rangle,
\quad \psi_{2L}(0,\sigma) |D, N\rangle = \psi^{\dagger}_{1R}(0,\sigma) |D,N\rangle, \\
\psi^{\dagger}_{1L}(0,\sigma) |D, N\rangle &=& -
\psi_{2R}(0,\sigma)|D,N\rangle, \quad \psi^{\dagger}_{2L}(0,\sigma)
|D, N\rangle = - \psi_{1R}(0,\sigma) |D,N\rangle. \nn \eeq
The boundary state $|D,N\rangle$ is
\begin{subequations}
\label{generallabel}
\begin{eqnarray}
&&|D,N\rangle_{NS} =
\prod_{r=\half}^\infty\left(1-\tilde\psi^{\dagger}_{1,-r}\psi^{\dagger}_{2,-r}
\right) \left(1+\psi^{\dagger}_{1,-r}\tilde\psi^{\dagger}_{2,-r}
\right) \left(1+\tilde \psi_{1,-r}\psi_{2,-r}\right) \left(1 -
\psi_{1,-r}\tilde \psi_{2,-r} \right) |0\rangle,
\\
&&|D,N\rangle_{R} =
\prod_{n=0}^\infty\left(1-\tilde\psi^{\dagger}_{1,-n}\psi^{\dagger}_{2,-n}
\right) \left(1+ \psi^{\dagger}_{1,-n}
\tilde\psi^{\dagger}_{2,-n}\right) \left(1+\tilde
\psi_{1,-n}\psi_{2,-n}\right) \left(1 -\psi_{1,-n}\tilde
\psi_{2,-n} \right) |----\rangle.
\end{eqnarray}
\end{subequations}

Finally we construct the boundary state $|D,D\rangle$.  It obeys
the boundary conditions for boson variables as follows
\beq X_L(0,\sigma)
|D,D\rangle = -X_R(0,\sigma) |D,D\rangle, \quad
Y_L(0,\sigma) |D,D\rangle = -Y_R(0,\sigma)
|D,D\rangle.
\eeq
The corresponding boundary condition in the fermion theory is
\beq
\psi_{1L}(0,\sigma) |D,D\rangle &=& i \psi_{1R}(0,\sigma)|D,D\rangle, \quad
\psi_{2L}(0,\sigma)|D,D\rangle = -i \psi_{2R}(0,\sigma)|D,D\rangle, \\
\psi^{\dagger}_{1L}(0,\sigma)|D,D\rangle &=&
i\psi^{\dagger}_{1R}(0,\sigma)|D,D\rangle, \quad
\psi^{\dagger}_{2L}(0,\sigma) |D,D\rangle = -i
\psi^{\dagger}_{2R}(0,\sigma)|D,D\rangle. \nn \eeq 
A solution of these equations is the boundary state
\begin{subequations}
\label{generallabel}
\beq
&&|D,D\rangle_{NS} = \prod_{r=\half}^\infty
\left(1+i\psi^{\dagger}_{1,-r}\tilde\psi_{1,-r}\right)
\left(1+i\psi_{1,-r}\tilde\psi^{\dagger}_{1,-r}\right)
\left(1-i\psi^{\dagger}_{2,-r}\tilde\psi_{2,-r}\right)
\left(1-i\psi_{2,-r}\tilde\psi^{\dagger}_{2,-r}\right) |0\rangle, 
\\
&&|D,D\rangle_{R} = -i\prod_{n=0}^\infty
\left(1+i\psi^{\dagger}_{1,-n}\tilde\psi_{1,-n}\right)
\left(1+i\psi_{1,-n}\tilde\psi^{\dagger}_{1,-n}\right)
\left(1-i\psi^{\dagger}_{2,-n}\tilde\psi_{2,-n}\right) \nn\\
&& \qquad \qquad \qquad \qquad  \qquad \qquad \qquad \qquad \qquad \qquad \qquad \qquad 
\qquad \qquad 
\left(1-i\psi_{2,-n}\tilde\psi^{\dagger}_{2,-n}\right) 
|-+-+\rangle.
\eeq
\end{subequations}
The boundary states that we have constructed are related to each other by
global $SU(2)_L\times SU(2)_L'$ rotations.  The easiest way to see this is
to note that the boundary conditions which they obey are related in this way.
It is easy to check that
 \beq\label{bstrans}
 |D,D\rangle= e^{-\pi i(J^1_0+ J^{\prime 1}_0)}|N,N\rangle
 = e^{-\pi i J^1_0}|N,D\rangle
 = e^{-\pi i J^{\prime 1}_0}|D,N\rangle.
 \eeq
Note that, unlike the transformation which obtained the rolling
tachyon boundary state in Eq.~(\ref{solution}), which was not
unitary, the transformations in Eq.~(\ref{bstrans}) are unitary
rotations.

 \vskip .75cm \begin{subsection}
 {\bf Open string partition functions }
 \end{subsection}

 In this subsection, we will consider the open string partition function
 for the boundary conditions described by the boundary state
 $|N,D\rangle$.
 We will show that the partition functions computed using the boson and the
 fermion representation of the boundary states are indeed identical.

 The boundary state in boson variables is
 \beq |N,D\rangle = \sum_{\pi_{1L},\pi_{1R}} \prod_{n=1}^\infty \exp
 \left(-\frac{1}{n}\left(\beta_{1,-n} \tbe_{2,-n} +
 \beta_{2,-n}
 \tbe_{1,-n}\right)\right)|\pi_{1L},\pi_{1R},-\pi_{1R},\,-\pi_{1L} \rangle. \label{bsnd}
 \eeq
Where $|\pi_{1L},\pi_{1R},
 \pi_{2L},\,\pi_{2R} \rangle=|\pi_{1L},\pi_{1R}>\otimes |
 \pi_{2L},\,\pi_{2R} \rangle$ is the direct product of vacuum states for
$\phi_1$ and $\phi_2$ and we have taken the boundary conditions into
account in constraining $\pi_{2L}=-\pi_{1R}$ and $\pi_{2R}=-\pi_{1L}$.
The momenta $\pi_{1L}$ and $\pi_{1R}$ in Eq.~(\ref{bsnd}) can have
either integer or half-odd-integer values and should be summed over
those values which are appropriate to the compactified $X$ and $Y$
bosons. The constraints were given in Eqs.~(\ref{221})-(\ref{224}).
For the states in $|N.D\rangle$, where $\pi_{2L}=\pi_{1R}$ and
$\pi_{2R}=\pi_{1L}$, these reduce to
\beq
\pi_{1L}-\pi_{1R}=w_X ~~,~~m_X=0~~,~~ w_Y=0
~~,~~\pi_{1L}+\pi_{1R}=m_Y.
\eeq
Here, $w_X$ and $m_Y$ are integers and there are two possibilities:
$\pi_{1L}$ and $\pi_{1R}$ are both integers (then $w_X$ and $m_Y$
are either both even or both odd), or $\pi_{1L}$ and $\pi_{1R}$ are
both half-odd-integers (and one of $m_X$ or $w_Y$ is even and the
other is odd).

 The open string partition function for the   $|N,D\rangle$ state is
 \begin{equation}\label{zndnd}
 Z_{(ND|ND)}[q] = \langle N,D| ~q^{ L_{10}+\tilde L_{10}+L_{20}+\tilde L_{20} } ~|N,D\rangle,
 \end{equation}
 where
 \begin{equation}
  L_{a0}+\tilde L_{a0} = \frac{1}{2}
 \left(\pi_{aL}^2+\pi_{aR}^2\right) +
 \sum_{n=1}^\infty
 \left(  \beta_{a,-n}
 \beta_{a,n}+ \tbe_{a,-n} \tbe_{a,n} \right)- \frac{1}{12},
 \end{equation}
labeled by $a=1,2$ belonging to $\phi_1$ or $\phi_2$.
 The partition function is computed as
 \begin{eqnarray}
 Z_{(ND|ND)}[q] &=& \sum_{\pi_{1L},\pi_{1R}} q^{\left[ \pi_{1L}^2 +
 \pi_{1R}^2- \frac{1}{6}\right]} \left(\prod_{n=1}
 \frac{1}{1-q^{2n}}\right)^2 \label{partnd1}
\nn\\
&=&\sum_{m,n\in{\cal Z}}\left( q^{  m^2 +
 n^2- \frac{1}{6} } + q^{  (m+\half)^2 +
 (n+\half)^2- \frac{1}{6} } \right)  \left(\prod_{n=1}^\infty
 \frac{1}{1-q^{2n}}\right)^2 \label{partnd2}
\end{eqnarray}
where we have explicitly indicated summations over the two possible
sets of values of $(\pi_{1L},\pi_{1R})$, where both are integers or
both are half-odd-integers.

To confirm that the above factorizes into partition functions for a
Neumann state and Dirichlet state, we would re-arrange the
summations as
\begin{eqnarray}
 Z_{(ND|ND)}[q]
&=&\sum_{m,n\in{\cal Z}}
q^{\half(m-n)^2-\frac{1}{12}}\left(q^{\half(m+n)^2-\frac{1}{12}}+q^{\half(m+n+1)^2-\frac{1}{12}}\right)
 \left(\prod_{n=1}^\infty
 \frac{1}{1-q^{2n}}\right)^2 \nn \\
 &=&\left(\sum_{m\in{\cal Z}}q^{\half
 m^2-\frac{1}{12}}\prod_{n=1}^\infty\frac{1}{1-q^{2n}}
 \right)^2=Z_{(N|N)}[q]Z_{(D|D)}[q]\label{bsquare}
\end{eqnarray}
where, in the last step, we observe that, if we keep $m-n$ fixed as
we vary $m$ and $n$, the values of $m+n$ and $m+n+1$ together sweep
over all of the integers. Because the $X$ and $Y$ bosons are
compactified at the self-dual radius, the Neumann and Dirichlet
partition functions are identical, thus the square in
Eq.~(\ref{bsquare}).

 We will now compute the open string partition function $Z_{(ND|ND)}[q]$ in the
 fermion theory.
 We must consider states both in the NS-NS and R-R sectors and project onto states
 where the total fermion number is even.
 In both R-R and NS-NS sectors, the total fermion number of the
boundary states is
 already even and the latter projection is not needed. Using the expression
 for the boundary state derived in the previous Subsection, we find:
 \begin{itemize}
 \item
 The partition function $Z_{(ND|ND)}[q]$ in the NS-NS sector:
 \beq
 Z^{NS}_{(ND|ND)}[q] &=& q^{-\frac{1}{6}} \prod_{n=\half}^\infty \left(1+
 q^{2n}\right)^4 \nn\\
 &=& \sum_{m,n} q^{\left[(m^2+n^2)
 -\frac{1}{6}\right]}\left(\prod_{n=1}^\infty\frac{1}{1-q^{2n}}\right)^2.
 \label{ndpf3}
 \eeq
 \item
 The partition function  $Z_{(ND|ND)}[q]$ in the R-R sector:
 \beq Z^R_{(ND|ND)}[q]
 &=& 2^2 q^{\frac{1}{3}} \prod_{n=1}^\infty\left(1+q^{2n}\right)^4
 \nn\\
 &=& \sum_{m,n} q^{\left[\left((m+\half)^2+
 (n+\half)^2\right)- \frac{1}{6}\right]}\left(\prod_{n=1}^\infty\frac{1}{
 1-q^{2 n}}\right)^2. \label{ndpf4}\eeq
 \end{itemize}
 The total partition function is given
 as a sum of contributions from both sectors, $ Z_{(ND|ND)}[q] =
 Z^{NS}_{(ND|ND)}[q] + Z^{R}_{(ND|ND)}[q]
 $.
 Upon combining the  expressions in (\ref{ndpf3}) and
 (\ref{ndpf4}) we find precise agreement with the partition function in the
 boson theory Eq.~(\ref{partnd2}).

 \vskip .75cm \begin{subsection}
 {\bf Exact Boundary State for the Rolling Tachyon}
 \end{subsection}

 Recall that the boundary state for the rolling tachyon is given in
 Eq.~(\ref{solution}), which we copy here:
 \beq
 |B,D \rangle = \exp\left(g\pi i J^+_0\right)|N, D\rangle
 ~~,~~
 J^+_0 = \int \frac{d\s}{2\pi}
 \psi^{\dagger}_{1L} \psi_{2L} = \sum_n \psi^{\dagger}_{1,n}
 \psi_{2,-n}
 \eeq
 with $n \in{\cal Z}+\half$ in the NS-NS sector,
 $n \in{\cal  Z}$ in the R-R sector. Combining (\ref{solution}) with
Eqs.~(\ref{bsnd1}) and (\ref{bsnd2}),
 we find the explicit form of the boundary  state is
\begin{subequations}
\label{generallabel}
\begin{eqnarray}
|B,D\rangle_{NS}&=& \prod_{r=\half}^\infty \left(1+\psi^{\dagger}_{1,-r}
\tpsi_{2,-r}\right)\left(1 + \tpsi^{\dagger}_{2,-r}
 \psi_{1,-r}-i\pi g
 \tpsi^{\dagger}_{2,-r} \psi_{2,-r} \right)\cdot
  \nonumber \\
 && \qquad \cdot\left(1 - \psi^{\dagger}_{2,-r}\tpsi_{1,-r}-i\pi g
 \psi^{\dagger}_{1,-r}\tilde\psi_{1,-r}\right)\left(1 -
 \tpsi^{\dagger}_{1,-r} \psi_{2,-r}\right)|0\rangle
 \label{bsbd1} \\
|B,D\rangle_{R} &=&-i\prod_{n=0}^\infty \left(1+\psi^{\dagger}_{1,-n}
\tpsi_{2,-n}\right)\left(1 + \tpsi^{\dagger}_{2,-n}
 \psi_{1,-n}-i\pi g
 \tpsi^{\dagger}_{2,-n} \psi_{2,-n} \right)\cdot
  \nonumber \\
 && \qquad \cdot\left(1 - \psi^{\dagger}_{2,-n}\tpsi_{1,-n}-i\pi g
 \psi^{\dagger}_{1,-n}\tilde\psi_{1,-n}\right)\left(1 -
 \tpsi^{\dagger}_{1,-n} \psi_{2,-n}\right)
|-+-+\rangle . \label{bsbd2}
\end{eqnarray}
\end{subequations}

The formulae presented in Eqs.~(\ref{bsbd1}) and (\ref{bsbd2}) are
our main result, the fermion representation of the boundary state of
the rolling tachyon.  Note that, in these states, the total fermion
number is even, so they  both correspond directly to boson states,
with no further need of projection.

\vskip .75cm

\section{Open string partition function for the Rolling Tachyon}

Now that we have an explicit expression for the boundary state in
Eqs.~(\ref{bsbd1}) and (\ref{bsbd2}) we can form the open string
partition function \footnote{For this computation we use the
identity
$\left(1+\psi_1^\dagger\tilde\psi_2\right)=e^{\psi_1^{\dagger}
\tilde\psi_2}$ to express the non-zero mode parts of the boundary
states using exponential operators operating on the vacuum.  Then
we use the identity $\langle0|e^{\psi^\dagger
M\tilde\psi}e^{\tilde\psi^\dagger N\tilde\psi}|0\rangle=\det(1+MN)$.}
\begin{equation}
Z_{(BD|BD)}[q]=\langle B,D| q^{L_0+\tilde L_0}|B,D\rangle.
\end{equation}
In the NS-NS sector this is given by
\begin{eqnarray}
Z_{(BD|BD)}^{NS}[q] &=& q^{-\frac{1}{6}}\prod_{r=\half}^\infty\left[(1+\zeta
q^{2r})(1+\zeta^{-1} q^{2r})\right]^2 \nonumber\\
&=& \left[ q^{-\frac{1}{12}}\sum_{n\in{\cal Z}}\zeta^n q^{
n^2}\prod_{n=1}^\infty\frac{1}{1-q^{2n}}\right]^2 \label{bigfor1} \\
&=&\left[ \sum_{m,n\in{\cal Z}}q^{\frac{1}{2}(m-n)^2}\zeta^{(m+n)}
q^{\frac{1}{2} (m+n)^2}\right]\left[q^{-\frac{1}{12}}
\prod_{n=1}^\infty\frac{1}{1-q^{2n}}\right]^2.\nn
\end{eqnarray}
where
$$
\zeta=1+\half\pi^2 g^2+g\pi\sqrt{1+\frac{1}{4}\pi^2g^2}
$$
and in the R-R sector,

\begin{eqnarray}
Z_{(BD|BD)}^{R}[q] &=& \left(\zeta^\half+\zeta^{-\half}\right)^2
q^{\frac{1}{3}} \prod_{m=1}^\infty\left[(1+\zeta
q^{2m})(1+\zeta^{-1} q^{2m})\right]^2 \nn \\ 
&=& \left[
q^{-\frac{1}{12}} \sum_{n\in{\cal Z}} \zeta^{n+\half} q^{
(n+\half)^2}
\prod_{n=1}^\infty\frac{1}{1-q^{2n}}\right]^2 \label{bigfor2} \\
& =&\left[ \sum_{m,n\in{\cal Z}}q^{\frac{1}{2}(m-n)^2}\zeta^{(m+n+1)}
q^{\frac{1}{2} (m+n+1)^2}\right]\left[q^{-\frac{1}{12}}
\prod_{n=1}^\infty\frac{1}{1-q^{2n}}\right]^2. \nn
\end{eqnarray}
It is easy to see that combining (\ref{bigfor1}) and (\ref{bigfor2})
obtains a summation over all integers $m-n$  where, holding $m-n$
fixed, $m+n$ and $m+n+1$ together also take on all integer values.
Thus, we get
\beq
Z_{(BD|BD)}^{NS}[q]+Z_{(BD|BD)}^{R}[q]= Z_{DD}[q]\cdot\left[ \sum_{
n\in{\cal Z}} \zeta^{n} q^{\frac{1}{2}
n^2}\right]\left[q^{-\frac{1}{12}}
\prod_{n=1}^\infty\frac{1}{1-q^{2n}}\right]
\eeq
where $Z_{DD}[q]$ is the Dirichlet partition function of the
$Y$-boson and our result for the rolling tachyon partition function
is
\begin{equation}
Z_{BB}[q]=q^{-\frac{1}{12}}
\sum_{ n\in{\cal Z}} \zeta^{n}
q^{\frac{1}{2} n^2}
  \prod_{n=1}^\infty\frac{1}{1-q^{2n}}.
\end{equation}

 \vskip .75cm \section{Disc Amplitude and Time Evolution of the Rolling Tachyon}

We have found a fermion representation of the states of compact
bosons.  However, the original problem of the rolling tachyon that
we set out to solve was in Minkowski space.  The rotation to
euclidean space should not produce a   compact boson. Instead, our
computations should be regarded as being incomplete, we should
discard all wrapped states and we have only found a subset of the
continuous spectrum of momentum states of the non-compact boson.
We might hope, nevertheless, that the discrete sampling of
momentum states that we have found are enough to determine the
spectrum.

For example, recall that the momentum and wrappping integers for
the X and Y-bosons are given by $(m_X,w_X,m_Y,w_Y)$ which in turn
are the combinations of the momenta
$(\pi_{1L},\pi_{1R},\pi_{2L},\pi_{2R})$ .given in
Eqs.~(\ref{221})-(\ref{224}).  To set the two wrapping numbers to
zero, we set $\pi_{1L}=\pi_{1R}\equiv\pi_1$ and
$\pi_{2L}=\pi_{2R}\equiv\pi_2$.  Then
$(m_X,w_X,m_Y,w_Y)=(\pi_1+\pi_2,0,\pi_1-\pi_2,0)$.

Consider the matrix element of this state with the $|N,D\rangle$
boundary state as given in the bosonic theory (\ref{bsnd}).
\begin{equation}
\langle\pi_1,\pi_1,\pi_2,\pi_2|N,D\rangle=\delta_{\pi_1+\pi_2} \end{equation}
If we define the position space state,
\beq
|X,Y\rangle=\frac{1}{2\pi}
\sum_{\pi_{1},\pi_2}e^{ -iX(\pi_1+\pi_2) - iY(\pi_1-\pi_2) }
|\pi_{1},\pi_{1},\pi_{2},\pi_{2}\rangle
\eeq
we see that
\beq
\langle X,Y|N,D\rangle=
\sum_{\pi_1,\pi_2}e^{iX(\pi_1+\pi_2)+iY(\pi_1-\pi_2)}\delta_{\pi_1,-\pi_2}
=\sum_{n}\delta(Y+2\pi n)
\eeq
which is what we would expect, given the Dirichlet boundary
condition for $Y$.  Note that this formula retains the information
that we treated $Y$ as a compact boson.

In the following, we will learn how to obtain the same result in
the fermion representation.  Then, we will compute $\langle X,Y|B,D\rangle$
using the fermion theory and compare it with the known result of
the current algebra computation (\ref{scalarf}).

\subsection{Momentum states}

 To begin, we must find the fermion representation of the
boson states which are oscillator vacua and eigenstates of the
momentum operators, $|\pi_{1L},\pi_{1R},\pi_{2L},\pi_{2R}\rangle$. In the
fermion representation in Eqs.~(\ref{currents1}) and
(\ref{currents2}),
\begin{subequations}
\label{generallabel}
\beq
\pi_{1L}&=&\oint
\frac{d\sigma}{2\pi}:\psi^{\dagger}_{1L}(0,\sigma)\psi_{1L}(0,\sigma):
~~,~~
\pi_{1R}=-\oint
\frac{d\sigma}{2\pi}:\psi^{\dagger}_{1R}(0,\sigma)\psi_{1R}(0,\sigma): 
\\
\pi_{2L}&=&-\oint
\frac{d\sigma}{2\pi}:\psi^{\dagger}_{2L}(0,\sigma)\psi_{2L}(0,\sigma):
~~,~~\pi_{2R}=\oint
\frac{d\sigma}{2\pi}:\psi^{\dagger}_{2R}(0,\sigma)\psi_{2R}(0,\sigma):.
\eeq
\end{subequations}
The state $|\pi_{1L},\pi_{1R},\pi_{2L},\pi_{2R}\rangle$ is an eigenstate
of fermion number operators and, since it is a vacuum for the boson
oscillators, it must be annihilated by the positively moded boson
oscillators.  These correspond to the positively moded components of
the fermion charge densities, which can also be found from
Eqs.~(\ref{currents1}) and (\ref{currents2}),
\begin{subequations}
\label{generallabel}
\beq
\beta_{1,n}&=&\sum_{r
 }\psi^{\dagger}_{1,n-r}\psi_{1,r}~,~
 \tilde\beta_{1,n}=-\sum_{r
 }\tilde\psi^{\dagger}_{1,n-r}\tilde\psi_{1,r}~,~ \\
 \beta_{2,n} &=& -\sum_{r
} \psi^{\dagger}_{2,n-r} \psi_{2,r} ~, ~ \tilde\beta_{2,n}=\sum_{r
}\tilde\psi^{\dagger}_{2,n-r}\tilde\psi_{2,r} ~,~n>0 .\eeq
\end{subequations}

For example, the unique eigenstate of $L_0+\tilde L_0$ with the
lowest eigenvalue is the state $|0,0,0,0\rangle$ of the bosons which is
annihilated by all positively moded boson oscillators.  This state
coincides with the vacuum of the fermions in the NS-NS sector, $|0\rangle$
which is annihilated by all positively moded fermion oscillators.

States with nonzero integer momenta are obtained by creating
fermions to fill up all fermion states up to a specific fermi level,
which determines the fermion number of the state and therefore its
momentum. For example, the state where $\pi_{1L}$ is a positive
integer is given by (up to a phase)
\begin{eqnarray}
 |\pi_{1L},0,0,0\rangle ~\sim~ \psi^{\dagger}_{1,-\pi_{1L}
+\half}\ldots\psi^{\dagger}_{1,-\frac{3}{2}}\psi^{\dagger}_{1,-\half}~|0\rangle
\nn\end{eqnarray} 
If $\pi_{1L}$ is a negative, rather than positive
integer, the state is
\begin{eqnarray}
 |\pi_{1L},0,0,0\rangle ~\sim~ \psi_{1,\pi_{1L}
+\half}\ldots\psi_{1,-\frac{3}{2}}\psi_{1,-\half}~|0\rangle
\nn\end{eqnarray}
These states satisfy the conditions
\begin{subequations}
\label{generallabel}
\begin{eqnarray}
\sum_{r=\half}^\infty\left(\psi^{\dagger}_{1,-r}\psi_{1,r}-
\psi_{1,-r}\psi^{\dagger}_{1,r}\right)|\pi_{1L},0,0,0\rangle &=& \pi_{1L}|\pi_{1L},0,0,0\rangle
\\
\sum_{r=-\infty}^\infty
\psi^{\dagger}_{1,n-r}\psi_{1,r}|\pi_{1L},0,0,0\rangle &=& 0~~~n>0
\end{eqnarray}
\end{subequations}
and thus coincide with the requisite states in the boson theory, up
to a phase which is not determined by these equations.

In the boson representation, in the phase convention that we have
chosen for the boundary state in Eqn.(\ref{bsnd}), the inner
products of momentum eigenstates and boundary states are all
either one or zero. We have found a phase convention which
produces this result in the fermion representation for the
momentum states of interest to us:
\begin{eqnarray}
  \begin{cases}
  \langle0|[ \psi_{1,\half}...\psi_{1,\pi_{1}-\half}]
  [ \tilde\psi^{\dagger}_{1,\pi_{1}-\half} ...
  \tilde\psi^{\dagger}_{1,\half}][ \psi^\dagger_{2,\half}    ...
  \psi^\dagger_{2,\pi_{2}-\half}][ \tilde\psi_{2,\pi_{2}-\half}  ...
  \tilde\psi_{2,\half}]~i^{\pi_1-\pi_2}~~  & {\pi_1>0,\pi_2>0}, \\
  \langle0|[ \psi_{1,\half}...\psi_{1,\pi_{1}-\half}]
  [ \tilde\psi^{\dagger}_{1,\pi_{1}-\half} ...
  \tilde\psi^{\dagger}_{1,\half}][ \psi_{2,\half}    ...
  \psi_{2,-\pi_{2}-\half}][ \tilde\psi^\dagger_{2,-\pi_{2}-\half}  ...
  \tilde\psi^\dagger_{2,\half}]~ i^{\pi_1+\pi_2} & 
  {\pi_1>0,\pi_2<0} \\
  \langle0|[\psi^{\dagger}_{1,\half}...\psi^{\dagger}_{1,-\pi_{1}-\half}]
  [ \tilde\psi_{1,-\pi_{1}-\half}...\tilde\psi_{1,\half}][ \psi_{2,\half}...
  \psi_{2,-\pi_{2}-\half}][ \tilde\psi^\dagger_{2,-\pi_{2}-\half}  ...
  \tilde\psi^\dagger_{2,\half}]~i^{-\pi_1+\pi_2} & 
  \pi_1<0,\pi_2<0 \\
  \langle0|[\psi^{\dagger}_{1,\half}...\psi^{\dagger}_{1,-\pi_{1}-\half}]
  [ \tilde\psi_{1,-\pi_{1}-\half}...
  \tilde\psi_{1,\half}][ \psi^\dagger_{2,\half}...
  \psi^\dagger_{2,\pi_{2}-\half}][ \tilde\psi_{2,\pi_{2}-\half}  ...
  \tilde\psi_{2,\half}]~i^{-\pi_1-\pi_2} &
  \pi_1<0,\pi_2>0
\end{cases}
\end{eqnarray}

 The additional phases which occur with each state are necessary
 to make the matrix elements of each state with the simply
 fermionic boundary states $|N,N\rangle,|D,N\rangle,|N,D\rangle,|D,D\rangle$ match those
 of the bosonic theory.  In the boson theory, the overlap of any
 momentum eigenstate with the simple boundary states is either
 zero or one, depending on the momenta.

The half-odd-integer momenta are found in the R-R sector.  There,
the choice of degenerate fermion vacuum is determined by the
requirement that the state be a boson oscillator vacuum.  For
example, the state with $\pi_{1L}>0$ is
$$
\psi^\dagger_{1,-\pi_{1L}+\half}...\psi^{\dagger}_{1,-2}\psi^{\dagger}_{1,-1}|+\rangle
$$
whereas, with $\pi_{1L}<0$,
$$
\psi_{1,\pi_{1L}+\half}...\psi_{1,-2}\psi_{1,-1}|-\rangle
$$
Again, our phase conventions can be summarized in bra-vector form
below
\begin{eqnarray}
  \begin{cases}
    \langle+--+|[ \psi_{1, 1}    ...
    \psi_{1,\pi_{1}-\half}]
    [ \tilde\psi^{\dagger}_{1,\pi_{1}-\half} ...
    \tilde\psi^{\dagger}_{1,1}] ~ [ \psi^\dagger_{2,1}  ...
    \psi^\dagger_{2,\pi_{2}-\half}][ \tilde\psi_{2,\pi_{2}-\half}  ...
    \tilde\psi_{2,1}]~ i^{\pi_1-\pi_2}  & \pi_1>0,\pi_2>0 \\
    \langle+-+-|[ \psi_{1, 1}  ...
    \psi_{1,\pi_{1}-\half}]
    [ \tilde\psi^{\dagger}_{1,\pi_{1}-\half} ...
    \tilde\psi^{\dagger}_{1,1}] [ \psi_{2,1}   ...
    \psi_{2,-\pi_{2}-\half}][ \tilde\psi^\dagger_{2,-\pi_{2}-\half}  ...
    \tilde\psi^\dagger_{2,1}]~i^{\pi_1+\pi_2+1} & 
    \pi_1>0,\pi_1<0 \\
    \langle-+-+|  [\psi^{\dagger}_{1,1}   ...
    \psi^{\dagger}_{1,-\pi_{1}-\half}]
    [ \tilde\psi_{1,-\pi_{1}-\half}  ...
    \tilde\psi_{1,1}] [ \psi^\dagger_{2,1}   ...
    \psi^\dagger_{2,\pi_{2}-\half}][ \tilde\psi_{2,\pi_{2}-\half}  ...
    \tilde\psi_{2,1}]~ i^{-\pi_{1}-\pi_2+1} & \pi_1<0, \pi_1>0 \\
    \langle-++-|  [\psi^{\dagger}_{1,1}   ...
    \psi^{\dagger}_{1,-\pi_{1}-\half}]
    [ \tilde\psi_{1,-\pi_{1}-\half}  ...
    \tilde\psi_{1,1}] [ \psi_{2,1}  ...
    \psi_{2,-\pi_{2}-\half}][ \tilde\psi^\dagger_{2,-\pi_{2}-\half}  ...
    \tilde\psi^\dagger_{2,1}]~ i^{-\pi_1+\pi_{2}+2} & \pi_1<0,\pi_2<0
  \end{cases}
\end{eqnarray}
Here, the ordering of the symbols in the vacuum bra-vector, for
example $\langle----|$  follows the same convention as for ket-vectors,
the first is the state of $\psi_{1,0}$, the second
$\tilde\psi_{1,0}$, etc., so that $\langle----|\psi_{1,0}=\langle+---|$.
Again, the phases of these states have been adjusted so that they
have the correct overlaps with the simple boundary states.

It is easy to check that, in both the NS and R sectors, the matrix
elements of our momentum eigenstates with boundary states are
\begin{eqnarray}
\langle\pi_{1},\pi_{1},\pi_{2},\pi_{2}|N,N\rangle &=&
\delta_{\pi_{1}}\delta_{\pi_{2}}
  \nn \\
\langle\pi_{1},\pi_{1},\pi_{2},\pi_{2}|N,D\rangle &=& \delta_{\pi_{1}+\pi_{2}}
  \nn \\
\langle\pi_{1},\pi_{1},\pi_{2},\pi_{2}|D,N\rangle &=& \delta_{\pi_{1}-\pi_{2}}
 \nn \\
\langle\pi_{1},\pi_{1},\pi_{2},\pi_{2}|D,D\rangle &=& 1
\end{eqnarray}
This agrees with what we would obtain in the boson theory.

\subsection{Rolling Tachyon}

Now, we are ready to compute the matrix elements of the momentum
eigenstates with the rolling tachyon state.
There are three cases of relative signs of the momenta. First,
when $\pi_{1},\pi_{2}\geq0$,
\begin{equation}
 \langle\pi_{1},\pi_{1},\pi_{2},\pi_{2}|B,D\rangle= (-\pi g)^{\pi_{1}}(-\pi
g)^{\pi_{2}}. \label{plusplus}
\end{equation}
Second, when $\pi_{1}\geq 0$ and $\pi_{2}\leq 0$ and
$\pi_{2}+\pi_{1}\geq0$,
\begin{equation}\label{plusminus}
  \langle\pi_{1},\pi_{1},\pi_{2},\pi_{2}|B,D\rangle=(-\pi
g)^{\pi_{1}+\pi_{2}}\delta_{\pi_{1}+\pi_{2}}.
\end{equation}
Third, when $\pi_{1}\leq 0$ and $\pi_{2}\geq 0$ and
$\pi_{2}+\pi_{1}\geq0$
\begin{equation}\label{minusplus}
 \langle\pi_{1},\pi_{1},\pi_{2},\pi_{2}|B,D\rangle=  (-\pi
g)^{\pi_{1}+\pi_{2}}\delta_{\pi_{1R}+\pi_{2L} }.
\end{equation}
Finally,
\begin{equation}
 \langle\pi_{1},\pi_{1},\pi_{2},\pi_{2}|B,D\rangle=0
\end{equation}
otherwise.

The above formulae apply to both the NS and R sectors where, in
the former the momenta are integers and in the latter they are
half-odd-integers.  It is then straightforward to find the disc
amplitude of the rolling tachyon.  We take the sums
\beq
\langle X,Y|B,D\rangle&=& \sum_{\pi_i,\pi_2}e^{iX(\pi_1+\pi_2)+iY(\pi_1-\pi_2)}
\langle\pi_1,\pi_1,\pi_2\pi_2|B,D\rangle \nn\\
&=&\sum_{\pi_1>0,\pi_2>0} \left( -\pi ge^{iX}\right)^{\pi_1+\pi_2}
e^{iY(\pi_1-\pi_2)}
+\sum_{\pi_1=0}^{-\infty}\sum_{\pi_2=-\pi_1}^\infty \left( -\pi g
e^{iX }\right)^{\pi_1+\pi_2}e^{iY(\pi_1-\pi_2)} \nn\\
& & \qquad +\sum_{\pi_2=0}^{-\infty}\sum_{\pi_1=-\pi_2}^\infty \left( -\pi g
e^{iX }\right)^{\pi_1+\pi_2}e^{iY(\pi_1-\pi_2)} \\
&=&\sum_{\pi_1+\pi_1= 0}^\infty\sum_{\pi_1-\pi_2=-\infty}^\infty
\left( -\pi ge^{iX}\right)^{\pi_1+\pi_2} e^{iY(\pi_1-\pi_2)}
=\frac{1}{1+\pi g e^{iX}}\sum_n\delta(Y+2\pi n). \nn
\eeq
Once we remove the amplitude for the Y-boson, and analytically
continue back to Euclidean space, $X\to -it$, we find that the
disc amplitude is
$$
\langle t|B\rangle=\frac{1}{1+\pi g e^t}
$$
which agrees with the result of current algebra, quoted in
(\ref{scalarf}).

 \vskip .75cm \section{Conclusion}   Time evolution of the unstable
D-brane, often termed as the rolling tachyon   in string theory,
is the key to understand the dynamics of the string theory.
Without understanding of this time dependent process the string
theory   should be regarded as incomplete. The most useful
framework to   study the time evolution of the unstable D-brane
may be the boundary state formulation, which describes most
efficiently the interaction   between the open string and a
D-brane. The tachyon mode of the open string  coordinate along the
time like direction of the target space  induces a nontrivial
interaction on the unstable D-brane. Upon taking the Wick rotation
on the string coordinate we find that the   system is be described
by a boundary conformal theory with a  marginal periodic boundary
interaction. The theory for the Wick rotated   system turns out to
be the boundary conformal theory which has a  wide range of
applications in condensed matter physics  \cite{call93,call94}
such as the Kondo model and junctions in quantum wires. Thanks to
the previous study of the system \cite{call93,call94} the exact
boundary state is known as a superposition of the $SU(2)$
Ishibashi states. However, the complete description of the time
evolution and the final fate of the unstable D-brane is still out
of reach of the current study since the transition functions
between the Ishibashi states and the perturbative string states
are not known in general except for a few
 states at low mass levels. Thus, there is an urgent need for a
more  suitable perturbative basis with which the exact boundary
state may be  written explicitly. This is the main purpose of the
present work.  

With the help of the extra boson field the entire
system of the unstable D-brane can be fermionized. The
advantage of the fermionization is that the boundary interaction can
be written as a bilinear operator, being a $SU(2)$ current operator,
in the fermionized theory. It readily implies that the boundary
condition to be  satisfied can be linear in terms of the fermion
fields. Taking advantage of  the fermionization we are able to
construct the exact boundary  state for the D-brane, which takes a
form of squeezed state in the  fermion theory. The fermionization or
bosonization of the conformal system with the marginal boundary
interaction is not new. Once the auxiliary boson field $Y$ is
introduced, the system  becomes the bosonized Kondo model
\cite{kondo,affleck} which has been extensively studied in
connection with various phenomena in condensed matter  physics. The
boundary conformal field theory of the rolling tachyon  in the
fermionized form is in fact identical to the Kondo model in its
original form. The fermionization of the same conformal model has
been also  considered by Polchinski and Thorlacius. But their study
was limited to the open string theory. 

In this Paper we develop a
closed string version of the fermionized conformal theory, which is
more suitable to discuss the rolling tachyon and construct the exact
boundary state.  Calculating the space-time dependent disk amplitude
with a scalar vertex explicitly we prove that the fermion
perturbative basis is the most suitable one  to investigate the time
evolution of the unstable D-brane at every level. In this work we
mainly deal with the simplest case of the  half-S-brane. Certainly,
the time evolution of the rolling tachyon with more  general tachyon
profiles can be discussed in this fermion representation developed
here. We may choose $|N,N\rangle$ as the boundary state to start
with to  study the S-brane within this framework. The present work
can be extended along various directions: The immediate one would be
the rolling tachyon with electric and magnetic fields \cite{rey}. It
is also interesting to apply the fermionization to the
supersymmetric rolling tachyon.  The outstanding problem is the
complete description of the time evolution of the rolling tachyon at
all higher levels and its final fate. \vskip .75cm

\section*{Acknowledgement}  This was supported in part by KOSEF
(Project No. R01-2000-00015) of Korea by NSERC of Canada. GWS
acknowledges the hospitality of the Kavli Institute for
Theoretical Physics where part of this work was done and the
partial support by the National Science Foundation under Grant
No.PHY99-07949. Part of TL's work was done at ICTP (Italy), KIAS (Korea),
PIMS and the Concentration in String Theory program of the Pacific
Institute for Mathematical Sciences and the Collaborative Research
Team on Strings and Particles of the Pacific Institute for
Theoretical Physics (Canada). Both authors acknowledge the
hospibality of the Banff International Research Station.  \vskip
.75cm
\section*{Appendix}

 \begin{center}   {\bf Cocycles for Fermion Operators}
\end{center}
 \vskip .75cm \begin{subsection}   {\bf One Dimensional System}
\end{subsection}   We parametrize the cocycles for the
fermions as follows   \beq  \psi_{L} &=& e^{- \frac{\pi i}{2}(\a^L
p_L + \b^L p_R)}
 e^{-\sqrt{2}i X_L}, \nn\\
 \psi_{R} &=& e^{\frac{\pi i}{2} (\a^R p_L + \b^R p_R)}
e^{\sqrt{2}i X_R}. \nn   \eeq   Their conjugates are   \beq
\psi_{L}^\dagger &=& e^{-\frac{\pi i}{2} \a^L}   e^{\frac{\pi
i}{2}(\a^L p_L + \b^L
 p_R)} e^{\sqrt{2}i X_L}, \nn\\
 \psi_{R}^\dagger &=& e^{-\frac{\pi i}{2} \b^R}
 e^{-\frac{\pi i}{2} (\a^R p_L + \b^R   p_R)} e^{-\sqrt{2}i
X_R}, \nn \eeq  where we make use of  \beq  [p_L, X_L] = [p_R, X_R]
= - i. \nn  \eeq  Here the left and right-moving bosons are
operators defined by the  mode expansion:
\begin{eqnarray}  X_L(\tau+i\sigma)~=~
\frac{x_L}{\sqrt{2}}-\frac{i}{\sqrt{2}}p_L(\tau+i\sigma)+
  \frac{i}{\sqrt{2}}\sum_{n\neq 0}\frac{1}{n} \alpha_n e^{-n(\tau+i\sigma)}   \nn \\
  X_R(\tau-i\sigma)~=~  \frac{x_R}{\sqrt{2}}-
\frac{i}{\sqrt{2}} p_R(\tau-i\sigma)+ \frac{i}{\sqrt{2}}\sum_{n\neq
0}\frac{1}{n} \tilde\alpha_n   e^{-n(\tau-i\sigma)} \nn
\end{eqnarray}  If the fermion field $\psi_L$ anti-commutes
with the fermion field  $\psi_R$, all anti-commutation relations
between the fermion  operators are satisfied. It yields the
following condition  \beq  e^{\frac{\pi i}{2}(\b^L-\a^R)} = -1. \eeq
  Thus, we need to impose   \beq \label{single1}
\b^L-\a^R = 2(2n+1), \quad n\in Z.   \eeq   The Neumann state
$|N\rangle$ is defined as   \beq  X_L(i\s)|N\rangle =
X_R(-i\s)|N\rangle,  \eeq   which reads in terms of normal mode
operators as  \beq   x_L|N\rangle = x_R|N\rangle, \quad p_L|N\rangle
=  -p_R|N\rangle,\quad \a_n|N\rangle = -\tal_{-n}|N\rangle \nn \eeq
  If we choose the cocycles appropriately we can simply realize
the Neumann   state in terms of fermion operators. It requires \beq
\label{single2} \a^L - \a^R -\b^L + \b^R =0.   \eeq   Under this
condition Eq.~(\ref{single2}) the Neumann state can   be expressed
in terms of fermion fields as follows  \beq  \psi_L|N\rangle =
e^{\frac{\pi i}{2}\b^L} \psi^{\dagger}_R   |N\rangle,\quad
\psi_L^\dagger|N\rangle = e^{\frac{\pi i}{2}(-\a^L+ \b^L-\b^R)}
\psi_R|N\rangle.  \eeq   This condition is consistent only if  \beq
\label{single3}  e^{-\frac{\pi i}{2}(\a^L- 2\b^L + \b^R)} =-1. \eeq
 We also require that the Dirichlet state $|D\rangle$ takes
a  simple form in the fermion theory. The Dirichlet condition is
given in the bosonic theory as  \beq  X_L|D\rangle = -X_R|D\rangle,
 \eeq   which can be read in terms of normal modes as
\beq  x_L|D\rangle = -x_R|D\rangle, \quad p_L|D\rangle =
p_R|D\rangle,\quad \a_n|D\rangle = \tal_{-n}|D\rangle. \nn  \eeq
  Applying the fermion operators on the Dirichlet state, we see
that  the Dirichlet condition is expressed in terms of the fermion
 operators only (without a nontrivial cocycle) only if  \beq
\label{single4}  \a^L+\a^R+\b^L+\b^R=0.   \eeq   If this condition
Eq.~(\ref{single4}) holds, the Dirichlet boundary   condition in the
fermion theory is given by   \beq   \psi_L|D\rangle = e^{\frac{\pi
i}{2}(\a^L+ \a^R)} \psi_R|D\rangle, \quad \psi^{\dagger}_L|D\rangle
= e^{-\frac{\pi i}{2}(\a^L+ \b^L)}   \psi_R^\dagger|D\rangle. \eeq
  This condition is consistent only if  \beq
e^{\frac{\pi i}{2}(\a^R-\b^L)} = -1.  \eeq   Note that this
condition coincides with the  anti-commutation condition
Eq.~(\ref{single1}).

 The solutions to the
Eqs.~(\ref{single1},\ref{single2},\ref{single3},\ref{single4}) exist
but not unique   \beq  \a^L = n, \quad \b^L = 2m+1, \quad \a^R=
-2m-1,  \quad \b^R = -n,  \eeq   where $n,m \in Z$. The simplest
solution may be with $n=m=0$   \beq  \a^L=0, \quad \b^L= 1,\quad
\a^R = -1, \quad \b^R =0. \eeq  With this choice the fermion field
operators are defined as  \beq   \psi_{L} &=& e^{-\frac{\pi i}{2}
p_R} e^{-\sqrt{2}i X_L}, \quad
  \psi_{R} = e^{-\frac{\pi i}{2} p_L} e^{\sqrt{2}i X_R} \\
  \psi^{\dagger}_{L} &=& e^{\frac{\pi i}{2} p_R} e^{\sqrt{2}i X_L},
 \quad
  \psi^{\dagger}_{R} = e^{\frac{\pi i}{2} p_L} e^{-\sqrt{2}i
  X_R}.\nn
  \eeq
  Accordingly the Neumann boundary condition and the Dirichlet
 boundary condition are written as   \beq   \psi_L|N\rangle
&=& i \psi^{\dagger}_R |N\rangle,\quad
 \psi_L^\dagger|N\rangle = i \psi_R|N\rangle, \\
 \psi_L|D\rangle &=& -i \psi_R |D\rangle,\quad
\psi_L^\dagger|D\rangle = -i \psi^{\dagger}_R |D\rangle.\nn  \eeq
  \vskip .75cm \begin{subsection}   {\bf Two Dimensional System}
\end{subsection}   We may parametrize the cocycles for the
two dimensional system with two   Dirac fermions in general as
follows   \beq   \psi_{1L} &=& e^{-\frac{\pi i}{2} (\a^L_{1j} p^j_L
+ \b^L_{1j}
  p^j_R)} e^{-\sqrt{2}i\Phi^1_L}, \nn\\
 \psi_{2L} &=& e^{\frac{\pi i}{2} (\a^L_{2j} p^j_L + \b^L_{2j}
  p^j_R)} e^{\sqrt{2}i\Phi^2_L},\\
  \psi_{1R} &=& e^{\frac{\pi i}{2} (\a^R_{1j} p^j_L + \b^R_{1j}
  p^j_R)} e^{\sqrt{2}i\Phi^1_R},\nn\\
 \psi_{2R} &=& e^{-\frac{\pi i}{2} (\a^R_{2j} p^j_L + \b^R_{2j}
 p^j_R)} e^{-\sqrt{2}i\Phi^2_R}\nn   \eeq   where $j = 1, 2$.
 Their conjugates are  \beq   \psi_{1L}^\dagger &=&
e^{-\frac{\pi i}{2} \a^L_{11}} e^{\frac{\pi i}{2} (\a^L_{1j} p^j_L +
\b^L_{1j}
 p^j_R)} e^{\sqrt{2}i\Phi^1_L}, \nn\\
  \psi_{2L}^\dagger &=& e^{-\frac{\pi i}{2} \a^L_{22}}
  e^{-\frac{\pi i}{2} (\a^L_{2j} p^j_L + \b^L_{2j}
  p^j_R)} e^{-\sqrt{2}i\Phi^2_L}, \\
 \psi_{1R}^\dagger &=& e^{-\frac{\pi i}{2} \b^R_{11}}
e^{-\frac{\pi i}{2} (\a^R_{1j} p^j_L + \b^R_{1j}
 p^j_R)} e^{-\sqrt{2}i\Phi^1_R}, \nn\\
 \psi_{2R}^\dagger &=& e^{-\frac{\pi i}{2} \b^R_{22}}
e^{\frac{\pi i}{2} (\a^R_{2j} p^j_L + \b^R_{2j}   p^j_R)}
e^{\sqrt{2}i\Phi^2_R}, \nn  \eeq  where we make use of  \beq [p^i_L,
\Phi^j_L] = [p^i_R, \Phi^j_R] = -i \delta^{ij}. \eeq   Requiring the
anti-commutation relation between $\psi_{iL}$ and   $\psi_{jL}$ we
find \beq \label{rel1} e^{\frac{\pi i}{2}(\a^L_{ij}-\a^L_{ji})} =
-1, \,\,\,{\rm for}\,\,\, i\not=j.  \eeq   Likewise requiring the
anti-commutation relation between $\psi_{iR}$ and $\psi_{jR}$, we
get \beq \label{rel2}  e^{\frac{\pi i}{2}(\b^R_{ij}-\b^R_{ji})} =
-1, \,\,\,{\rm for}\,\,\, i\not=j. \eeq   The anti-commutation
relation between $\psi_{iL}$ and $\psi_{jR}$  yields the following
condition  \beq \label{rel3}  e^{\frac{\pi
i}{2}(\b^L_{ij}-\a^R_{ji})} = -1.  \eeq  Then anti-commutation
relations will be ensured between all  fermion operators.

The simple boundary conditions should be realized in terms of the
  fermion operators (without the cocycles). We begin with the
 boundary state $|N,N\rangle$.  The $|N,N\rangle$ boundary
condition is given in terms of the bosonic operator   as follows
  \beq   \Phi^i_L|N,N\rangle = \Phi^i_R|N,N\rangle   \eeq
  This condition can be realized simply in fermion theory if the
  following conditions are satisfied  \beq \label{cond1}
\a^L_{ij}-\a^R_{ij} -\b^L_{ij} +\b^R_{ij} = 0.  \eeq   Under this
condition we may write define $|N,N\rangle$ in the  fermion theory
as  \beq \psi^i_L|N,N\rangle = e^{\frac{\pi i}{2}\b^L_{ii}}
\psi^{i\dagger}_{R}|N,N\rangle, \quad  \psi^{i \dagger}_L|N,N\rangle
= e^{-\frac{\pi
i}{2}(\a^L_{ii}-\b^L_{ii}+\b^R_{ii})}\psi^i_R|N,N\rangle. \eeq The
consistency requires  \beq \label{cond2}   e^{\frac{\pi
i}{2}(-\a^L_{ii}+2 \b^L_{ii} -\b^R_{ii})}= -1.   \eeq

 In the two
dimensional system we have well defined fermion  current operators
$J^1$ and $J^{\prime 1}$   \beq  e^{-\pi i J^1_0} J^3 e^{\pi i
J^1_0} &=& e^{-\pi i J^1_0} i\partial_z X
 e^{\pi i J^1_0}= - i\partial_z X, \\
 e^{-\pi i J^{\prime 1}_0} J^{\prime 3} e^{\pi i J^{\prime 1}_0}
 &=& e^{-\pi i J^{\prime 1}_0} i\partial_z Y e^{\pi i J^{\prime
1}_0}  = - i\partial_z Y. \nn  \eeq  Using these fermion current
operators we can obtain the boundary  states $|N,D\rangle$
$|D,N\rangle$ and $|D,D\rangle$ consistently  if we once have the
boundary state $|N,N\rangle$ in the fermion  theory \beq |D,N\rangle
= e^{-\pi i J^1_0} |N,N\rangle, \quad  |N,D\rangle = e^{-\pi i
J^{\prime 1}_0} |N,N\rangle, \quad  |D,D\rangle = e^{- \pi i (J^1_0
+ J^{\prime 1}_0)}|N,N\rangle.  \eeq

 The tachyon boundary
interaction term may be written in terms of  fermion operators only
if we choose $\a's$ and $\b's$  appropriately. Since we may write
the tachyon interaction term  \beq  :e^{i(X_L+X_R)}: |B,D\rangle =
:e^{i(X_L+Y_L)}::e^{i(X_R+Y_R)}:|B,D\rangle,  \eeq  we find that it
can be rewritten in terms of fermion field operators only if the
following condition is satisfied  \beq \label{cond3}  \a^L_{1j} +
\a^R_{1j} =0, \quad  \b^L_{1j} + \b^R_{1j} =0,  \eeq  Then the
tachyon boundary interaction term can be written as  \beq
:e^{i(X_L+X_R)}: |B,D\rangle = e^{\frac{\pi i}{2}(\a^L_{11}+
\a^R_{11})} \psi^{\dagger}_{1L} \psi_{1R} |B,D\rangle =
\psi^{\dagger}_{1L}  \psi_{1R} |B,D\rangle.  \eeq

 We note that the
tachyon interaction term can be also written as  \beq
:e^{i(X_L+X_R)}: |B,D\rangle=
:e^{i(X_L-Y_L)}::e^{i(X_R-Y_R)}:|B,D\rangle.  \eeq  This equation
can be expressed in terms of the fermion field  operators if the
following condition is satisfied  \beq \label{cond4}  \a^L_{2j} +
\a^R_{2j} = 0, \quad  \b^L_{2j} + \b^R_{2j} = 0.  \eeq  Under this
condition Eq.~(\ref{cond4})  the tachyon boundary interaction term
can be written as  \beq  :e^{i(X_L+X_R)}: |B,D\rangle = e^{\frac{\pi
i}{2}(\b^L_{22}+   \b^R_{22})} \psi^{\dagger}_{2R} \psi_{2L}
|B,D\rangle = \psi^{\dagger}_{2R} \psi_{2L} |B,D\rangle.  \eeq

 It is
not difficult to show that there exist solutions to the  conditions
Eqs.~(\ref{rel1},\ref{rel2},\ref{rel3},\ref{cond1},\ref{cond2},\ref{cond3},\ref{cond4}).
 \beq
 \a^L_{11} &=& \b^L_{11} = -\a^R_{11} = -\b^R_{11} = 2n+1, \nn\\
 \a^L_{12} &=& \b^L_{12} = -\a^R_{12} = -\b^R_{12} = 2k+2l+2, \\
 \a^L_{21} &=& \b^L_{21} = -\a^R_{21} = -\b^R_{21} = 2k-2l, \nn\\
 \a^L_{22} &=& \b^L_{22} = -\a^R_{22} = -\b^R_{22} = 2m+1, \nn
\eeq  where $n, m, k, l \in Z$.  Choosing $n=m=k=l=0$, we may get
the simplest solution \beq \label{simplest}
 \a^L_{11} &=& \b^L_{11} = -\a^R_{11} = -\b^R_{11} = 1, \nn\\
 \a^L_{12} &=& \b^L_{12} = -\a^R_{12} = -\b^R_{12} = 2, \\
 \a^L_{21} &=& \b^L_{21} = -\a^R_{21} = -\b^R_{21} = 0, \nn\\
 \a^L_{22} &=& \b^L_{22} = -\a^R_{22} = -\b^R_{22} = 1, \nn
\eeq  With this solution the boundary condition for $|N,N\rangle$
reads  as  \beq  \psi^i_L|N,N\rangle = i
\psi^{i\dagger}_R|N,N\rangle, \quad  \psi^{i\dagger}_L|N,N\rangle =
i \psi^i_R|N,N\rangle, \quad  i = 1,2.  \eeq  Applying $e^{-\pi i
J^1_0}$ to $|N,N\rangle$ we obtain the  boundary state $|D,N\rangle
=e^{-\pi i J^1_0}|N,N\rangle$  \beq e^{-\pi i J^1_0} \psi^i_L e^{\pi
i J^1_0} |D,N\rangle =   i\psi^{i\dagger}_R|D,N\rangle.  \eeq Since
  \beq \label{trns1}  e^{-\pi i J^1_0} \psi^1_L e^{\pi i
J^1_0} &=& i \psi^2_L, \quad
 e^{-\pi i J^1_0} \psi^2_L e^{\pi i J^1_0} = i \psi^1_L, \\
 e^{-\pi i J^1_0} \psi^{1\dagger}_L e^{\pi i J^1_0} &=& -i
\psi^{2\dagger}_L, \quad  e^{-\pi i J^1_0} \psi^{2\dagger}_L e^{\pi
i J^1_0} = -i \psi^{1\dagger}_L, \nn  \eeq  we get the boundary
condition for $|D,N\rangle$ in the fermion  theory  \beq
\psi^1_L|D,N\rangle &=& \psi^{2\dagger}_R|D,N\rangle,\quad
 \psi^2_L|D,N\rangle = \psi^{1\dagger}_R|D,N\rangle \\
 \psi^{1\dagger}_L|D,N\rangle &=& -\psi^{2}_R|D,N\rangle,\quad
\psi^{2\dagger}_L|D,N\rangle = - \psi^{1}_R|D,N\rangle. \nn  \eeq

The boundary state $|N,D\rangle$ can be obtained by applying
$e^{-\pi i J^{\prime 1}_0}$ to $|N,N\rangle$  \beq  |N,D\rangle =
e^{-\pi i J^{\prime 1}_0} |N,N\rangle, \quad  J^{\prime 1}_0 =
\frac{1}{2} \int \frac{d\s}{2\pi} \left(  \psi^{\dagger}_{1L}
\psi^{\dagger}_{2L} + \psi_{2L} \psi_{1L}\right).  \eeq  Since  \beq
\label{trns2}  e^{-\pi i J^{\prime 1}_0} \psi^1_L e^{\pi i J^{\prime
1}_0} &=&  i \psi^{2\dagger}_L, \quad  e^{-\pi i J^{1\prime}_0}
\psi^{2}_L e^{\pi i J^{\prime 1}_0} =
 -i\psi^{1\dagger}_L, \\
 e^{-\pi i J^{1\prime}_0} \psi^{1\dagger}_L e^{\pi i J^{\prime
1}_0}  &=& -i \psi^{2}_L, \quad  e^{-\pi i J^{1\prime}_0}
\psi^{2\dagger}_L e^{\pi i J^{\prime 1}_0}  = i \psi^1_L, \nn  \eeq
 we get the boundary condition for $|N,D\rangle$ in the fermion
 theory  \beq  \psi^1_L|N,D\rangle &=&
\psi^{2}_R|N,D\rangle,\quad
 \psi^{2}_L|N,D\rangle = - \psi^{1}_R|N,D\rangle, \\
 \psi^{1\dagger}_L|N,D\rangle &=& -
\psi^{2\dagger}_R|N,D\rangle,\quad  \psi^{2\dagger}_L|N,D\rangle =
\psi^{1\dagger}_R|N,D\rangle \nn.  \eeq

 Finally applying $e^{-\pi i
(J^1_0+ J^{\prime 1}_0)}$ to  $|N,N\rangle$, we get the boundary
state $|D,D\rangle$ and its boundary condition. With the help of
Eq.~(\ref{trns1}) and  Eq.~(\ref{trns2})  \beq  \psi_{1L}
|D,D\rangle &=& i \psi_{1R}|D,D\rangle, \quad
 \psi_{2L}|D,D\rangle = -i \psi_{2R}|D,D\rangle, \\
 \psi^{\dagger}_{1L}|D,D\rangle &=& i\psi^{\dagger}_{1R}|D,D\rangle,
 \quad \psi^{\dagger}_{2L} |D,D\rangle = -i
\psi^{\dagger}_{2R}|D,D\rangle. \nn   \eeq

\end{document}